\documentclass[a4paper]{aa}

\usepackage{graphicx}
\usepackage{natbib}
\usepackage{txfonts}
\usepackage{color}

\newcommand{\invers}[1]{{\sc invers{#1}}}

\newcommand{\llm}[1]{{\sc LLmodels}}
\newcommand{\eq}[1]{{Eq.~{(\ref{#1})}}}

\newcommand{\kms}{km\,s$^{-1}$}

\newcommand{\fe}[1]{\ion{Fe}{#1}}
\newcommand{\nd}[1]{\ion{Nd}{#1}}
\newcommand{\sod}[1]{\ion{Na}{#1}}

\newcommand{\bfa}[1]{#1}

\graphicspath{{fig/}}

\begin{document}
\title{Three-dimensional magnetic and abundance mapping \\ of the cool Ap star HD\,24712}

\subtitle{II. \bfa{Two-dimensional} Magnetic Doppler Imaging in all four Stokes parameters%
\thanks{Based on observations collected at the European Southern Observatory, Chile (ESO programs 084.D-0338, 085.D-0296, 086.D-0240).}}

\author{N.~Rusomarov\inst{1}
\and O.~Kochukhov\inst{1}
   \and T.~Ryabchikova\inst{2}
   \and N.~Piskunov\inst{1}}

\institute{
Department of Physics and Astronomy, Uppsala University, Box 516, 75120 Uppsala, Sweden
\and
Institue of Astronomy, Russian Academy of Sciences, \bfa{Pyatnistkaya} 48, 119017 Moscow, Russia
}

\date{Received 9 July 2014/ Accepted 24 September 2014}

\titlerunning{Magnetic Doppler Imaging of HD\,24712 in all four Stokes parameters}
\authorrunning{N.~Rusomarov et al.}

\abstract
{}
{We present a magnetic Doppler imaging study from all Stokes parameters of the cool, chemically peculiar star HD\,24712. This is the very first such analysis performed at a resolving power exceeding $10^5$.}
{The analysis is performed on the basis of phase-resolved observations of line profiles in all four Stokes parameters obtained with the HARPSpol instrument attached at the 3.6-m ESO telescope. We use the magnetic Doppler imaging code, \invers{10}, which allows us to derive the magnetic field geometry and surface chemical abundance distributions simultaneously.}
{We report magnetic maps of HD\,24712 recovered from a selection of \fe{i}, \fe{ii}, \nd{iii}, and \sod{i} lines with strong polarization signals in all Stokes parameters. Our magnetic maps successfully reproduce most of the details available from our observation data. We used these magnetic field maps to produce abundance distribution map of Ca. This new analysis shows that the surface magnetic field of HD\,24712 has a dominant dipolar component with a weak contribution from higher-order harmonics. The surface abundance distributions of Fe and Ca show enhancements near the magnetic equator with an underabundant patch at the visible (positive) magnetic pole; Nd is highly abundant around the positive magnetic pole. The Na abundance map shows a high overabundance around the negative magnetic pole.}
{Based on our investigation and similar recent four Stokes parameter magnetic mapping studies, we present a tentative evidence for the hypothesis that Ap stars with dipole-like fields are older than stars with magnetic fields that have more small-scale structures. We find that our abundance maps are inconsistent with recent theoretical calculations of atomic diffusion in presence of magnetic fields.}

\keywords{stars: chemically peculiar -- stars: atmospheres -- stars: abundances -- stars: individual: HD\,24712 -- stars: magnetic field}

\maketitle

\section{Introduction}
\label{sec:intro}
The interplay between the magnetic field and chemical spots in magnetic Ap stars has been the subject of many investigations. Historically, these studies were limited from one hand by the instrumental capabilities at that time and from the other hand by the available data. As a result the majority of the magnetic field data in the literature are measurements of the mean longitudinal magnetic field from circular polarization within absorption lines. The polarization features in the spectra of magnetic Ap stars are produced by the Zeeman effect --- even a magnetic field with strength of tens of Gauss will cause spectral lines to split into multiple components and become polarized.

In spite of their usefulness these measurements have limited scientific value because they contain relatively little information about the topology of the magnetic field and can only constrain the strength and orientation of the dipolar component of the magnetic field. One of the important results that arouse from these studies was the oblique rotator model \citep{Stibbs1950p395}, which explains the observed periodic variability of the longitudinal magnetic field and the strength of the spectral lines as a function of time by proposing that magnetic Ap stars have a simple axisymmetric magnetic field inclined relative to the rotational axis and frozen into the rigidly rotating star.

As a way to extract more information about magnetic fields of Ap stars from their circular polarization measurements the so-called moments technique was proposed \citep{Mathys1988p179,Mathys97p475}. Other attempts were directed at detecting and interpreting net broad-band linear polarization (BBLP) produced by the Zeeman effect, which can constrain the mean transverse component of the magnetic field \citep{Landolfi1993p285,Leroy1995p79}. Both of these methods, ultimately, suffer from a number of severe limitations; the moments technique cannot account for the severe blending of spectral lines in spectra of Ap stars, has strong dependency on photon flux and on the number of lines available for analysis; the measurement of BBLP can only be done for stars for which interstellar polarization is small, have measurable signal, which depends on the spectral type and the geometry of the magnetic field. Additionally both methods do not account for the effects of the inhomogeneous distribution of chemical abundances over the stellar surface and have a limited capability to constrain higher-order multipolar components of the magnetic field \citep{Bagnulo1995p459,Landstreet2000p213}. Most importantly, modeling of these observables does not necessarily recover a field configuration that can reproduce Stokes parameter profiles in spectra of Ap stars \citep{Bagnulo2001p889}.

The advent of high-resolution spectropolarimeters available on medium size telescopes opened the possibility to acquire phase-resolved high-resolution spectropolarimetric observations in circular and linear polarization. \bfa{\citet{Wade00p823} published the first such observations of magnetic Ap and Bp stars, which proved that it is possible to detect linear polarization signatures in individual lines, and that these signals indeed are a result of the Zeeman effect. However, \citet{Wade00p823} could detect linear polarization signatures in only the most magnetically sensitive lines for the spectra with exceptionally high signal-to-noise ratio.}

Further progress in the study of magnetic fields and abundance structures in Ap stars was made with the introduction of magnetic Doppler imaging (MDI) \citep{Piskunov2002p736,Kochukhov2002p868}. This method is based on time-series of high-resolution Stokes profiles of spectral lines and allows one to derive maps of the magnetic field and abundance distribution of chemical elements on the stellar surface. It has been shown that MDI based on four Stokes parameter data does not require \textit{a priori} information about the global magnetic field topology, thus removing the biggest limitation of the traditional methods. Despite the obvious advantages of four Stokes parameter MDI over the traditional methods, there are only two such studies for magnetic Ap stars. The first investigation was performed for the star 53~Camelopardalis (HD\,65339) \citep{Kochukhov2004p613} for which previous studies using low-order multipole parametrization for the magnetic field \citep{Bagnulo2001p889} showed particularly strong disagreement between the predicted and observed Stokes~$Q$ and $U$ profiles. In the light of that discovery \citet{Kochukhov2004p613} showed that the magnetic field of 53~Cam has a toroidal component that is comparable in strength to the poloidal one but dominates on spatial scales of $30^\circ\text{--}\,40^\circ$ compared to the mostly dipolar nature of the poloidal component. The second MDI study \citep{Kochukhov10p13}, performed for the magnetic Ap star $\alpha^2$\,CVn~(HD\,112413) resulted in magnetic maps that revealed dipolar-like magnetic field only on the largest scales with definite asymmetry in the field strength between the positive and negative magnetic poles, and small-scale features for which the magnetic field strength is significantly higher than in the surrounding areas. \citet{Kochukhov10p13} in their analysis proved that these small-scale features are necessary for the reproduction of the observed polarization signatures. Recently, \citet{Silvester2014p182} derived updated magnetic field and abundance maps of $\alpha^2$\,CVn from new spectropolarimetric observations \citep{Silvester2012p1003} with superior quality relative to the data used in the initial study. The updated maps confirm the original results of \citet{Kochukhov10p13} providing observational evidence that the structure of the magnetic field of $\alpha^2$\,CVn is stable over the period of a decade, and that the reconstructed magnetic field is a realistic representation of the magnetic field of $\alpha^2$\,CVn.

In the context of these results it has become obvious that more such studies are necessary if we want to get more insight into the problem of magnetic fields of Ap stars. For this purpose we have started a new program aimed at observing Ap stars in all four Stokes parameters with the HARPSpol spectropolarimeter \citep{Piskunov11p7} at the ESO 3.6-m telescope. This instrument allows for full Stokes vector observations with spectral resolution greater than $10^5$. As our first object we have chosen HD\,24712, which is one of the coolest Ap stars showing both stratification of chemical elements and oblique rotator variations \citep{Ryabchikova97p1137,Lueftinger10p71,Shulyak09p879}.

The work presented here focuses primarily on deriving maps of the magnetic field and horizontal (surface) abundance distribution of several chemical elements from the four Stokes observations of HD\,24712. This is the first such analysis for a rapidly oscillating Ap (roAp) star. The previous MDI study of HD\,24712 was performed by \citet{Lueftinger10p71} and employed only Stokes~$I$ and $V$ parameters and multipolar regularization \citep[see][]{Piskunov2002p736} for the magnetic field.

The present paper is the second in a series of publications where we investigate the possibility of performing 3-D MDI analysis. We believe that by overcoming one of the limitations of our current MDI code \citep{Piskunov2002p736}---the inability to incorporate vertical abundance stratification of chemical elements---we can better understand the relation between chemical spots and magnetic field, and between horizontal and vertical abundance structures. This will allow us to use the excellent observational data to their full extent. Detailed introduction to the reasoning behind our attempts at 3-D MDI analysis can be found in our first paper \citep[herein refered as paper~I;][]{Rusomarov2013p8}. Because of the complexity and volume of such undertaking the 3-D analysis will be addressed in our next paper.

The paper is organized as follows: Sect.~\ref{sec:obs} briefly explains the observations, Sect.~\ref{sec:mdi} gives a short introduction to the principles of MDI, describes the spectral lines used in the inversion procedure, the choice of the model atmosphere, and the optimization of the global parameters. Sect.~\ref{sec:res} gives the results of the MDI analysis. Sect.~\ref{sec:dis} and \ref{sec:concl} contain discussion and conclusions.

\section{Spectropolarimetric observations}
\label{sec:obs}
HD\,24712 was observed during 2010--2011 over sixteen nights. In total we obtained 43 individual Stokes parameter observations. The observations have signal-to-noise ratio of 300--600 and resolving power exceeding $10^5$. The spectra were obtained with the HARPS spectrograph \citep{Mayor2003p20} in its polarimeter mode \citep{Snik2011p237,Piskunov11p7} at the ESO 3.6-m telescope at La Silla, Chile. The resulting spectra have excellent rotational phase coverage. We have thirteen Stokes~$IQUV$ observations, two Stokes~$IV$, and one Stokes~$IQU$ observation. 

Detailed analysis of the full Stokes vector spectropolarimetric data set of HD\,24712 can be found in paper~I. We redirect the reader to that paper for the detailed discussion of the observations, the data reduction, the description of the instrument and the observation procedure. An important finding of that study is the absence of significant spurious polarimetric signals that can affect our MDI analysis.

\section{Magnetic Doppler Imaging}
\label{sec:mdi}

\subsection{Methodology}
\label{sec:mdi:intro}
Magnetic Doppler imaging techniques using circular and linear polarization spectra are thoroughly discussed in the paper by \citet{Piskunov2002p736}. Here we give a short overview of some important principles of our MDI methodology and its implementation in the \invers{10} code used in this study.

The general idea behind MDI is to search for an optimal fit of synthetic spectra to a set of observational data by adjusting the surface distribution of magnetic field and abundances of chemical elements. Formally, this means that MDI can be considered as a least-squares minimization problem:
\begin{equation}
  \Psi = \sum_{k} \mathcal{D}_k + \mathcal{R} \rightarrow \mathrm{min},
\end{equation}
where $\mathcal{D}_k$ is the discrepancy between the observed and calculated spectra for a given Stokes parameter, and $\mathcal{R}$ is the regularization functional. In the case of HD\,24712 the index $k$ goes through the available Stokes parameters, $k = \{I,Q,U,V\}$. The discrepancy $\mathcal{D}_k$ for a given $k$ is
\begin{equation}
  \label{eq:mdi:Dk}
  D_k = \sum_{\varphi \lambda} w_k \left[ F_{k \varphi \lambda}^\mathrm{c}(\vec{B},\varepsilon^1, \varepsilon^2,\ldots ) - F_{k \varphi \lambda}^\mathrm{o}\right]^2 / \sigma^2_{k \varphi \lambda},
\end{equation}
where the summation goes through the given rotational phases $\varphi$ and wavelength points $\lambda$ of the computed $F_{k \varphi \lambda}^\mathrm{c}$ and observed $F_{k \varphi \lambda}^\mathrm{o}$ Stokes parameter profiles. The synthetic Stokes profiles $F_{k \varphi \lambda}^\mathrm{c}$ depend on the magnetic field $\vec{B}$ and abundance distributions $\varepsilon^1$, $\varepsilon^2$, etc. The weights $w_k$ are introduced to make sure that the relative contributions to the total discrepancy function $\Psi$ of different Stokes parameters are approximately the same. For the case of HD\,24712 the relative weights are $w_I\mathrm{:}w_Q\mathrm{:}w_U\mathrm{:}w_V = 1\mathrm{:}19\mathrm{:}16\mathrm{:}3.5$. The relative weights are automatically calculated from the phase-averaged amplitudes of the observed Stokes parameters.

The stellar surface is divided into $N$ approximately equal area zones. For the present study of HD\,24712 we use a grid consisting of 1176 surface elements. This grid is sufficient given the spectral resolution of the observational data and $v_e \sin i = 5.6 \pm 2.3$\,\kms{} \citep{Ryabchikova97p1137}. On this surface grid we define the magnetic field $\vec{B}$ and the abundance distributions $\varepsilon^1$, $\varepsilon^2$, etc. The synthetic Stokes profiles, $F_{k \varphi \lambda}^\mathrm{c}$, used in \eq{eq:mdi:Dk}, are computed by integrating the local Stokes profiles across the visible disk of the star for each phase $\varphi$ on the wavelength grid of the original spectra $F_{k \varphi \lambda}^\mathrm{o}$. This integration procedure also takes into account the projected area of each surface element for each $\varphi$. The local profiles before each integration are convolved with a Gaussian function to take into account the finite spectral resolution of the instrument, and are Doppler shifted for each rotational phase. At last, the integrated Stokes profiles are normalized by the phase-independent, unpolarized continuum.

The local Stokes profiles are computed for a given stellar model atmosphere and depend on the local values of the magnetic field vector and chemical abundances. In contrast to other MDI codes \citep{Donati1997p1135}, \invers{10} does not use simplifying approximations in the form of fixed local Gaussian profiles, or Milne-Eddington atmosphere, instead equations of polarized radiative transfer are solved numerically each time to derive the local Stokes profiles. 

Following \citet{Donati06p629} and \citet{Kochukhov2014p83} we use the spherical harmonics expansion of the magnetic field into a sum of poloidal and toroidal components. In this formalism one needs to find a set of spherical harmonic coefficients $\alpha_{\ell,m}$, $\beta_{\ell,m}$, and $\gamma_{\ell,m}$ that represent the radial poloidal, horizontal poloidal, and toroidal components respectively. This approach has many benefits: it is trivial to test specific magnetic field configurations, e.g., dipole, dipole plus quadrupole; the number of spherical harmonics coefficients is around five to ten times less than what would be required if we directly map the magnetic field components, which speeds up our calculations due to a smaller number of free parameters; the divergence free condition for the magnetic field is automatically satisfied in this formalism; we can easily calculate relative energies of the poloidal and toroidal components of the magnetic field according to the surface integrals $\int \vec{B}^2$.

In the case of HD\,24712 this formalism is the most appropriate due to the star showing only the positive magnetic pole to the observer. This prevents us from directly mapping the magnetic field components and applying Tikhonov regularization individually to the radial, meridional, and azimuthal maps as discussed by \citet{Piskunov2002p736}.

For a star with projected rotational velocity $v_e \sin i \simeq 5.6$\,\kms{} and full-width at half maximum of the intrinsic profile measured in the absence of rotation $W \simeq 3.2$\,\kms{} the number of resolved equatorial elements across the disk of the star will be approximately $2 v_e \sin i / W \simeq 4$, which implies that modes of the order of $\ell=8$ are resolvable. In our numerical experiments we used $\ell \le 10$ and found that for $\ell > 6$ all spherical harmonics coefficients had insignificant influence on the quality of the fit. Therefore we truncated the spherical harmonics expansion at $\ell_\mathrm{max} = 6$.

An important part of each Doppler imaging code is the regularization method, which is necessary for ensuring stability of the solution with respect to the initial guess, surface discretization and phase sampling of the observations. The general form of the regularization functional $\mathcal{R}$ implemented in \invers{10} for this study is
\begin{equation}
  \label{eq:mdi:regul}
  \mathcal{R} = \Lambda_a \mathcal{R}_a + \Lambda_f \mathcal{R}_f,
\end{equation}
where
\begin{equation}
  \label{eq:mdi:regul:ra}
  \mathcal{R}_a = \sum_i \sum_j \left[ (\varepsilon_i^1 - \varepsilon_j^1)^2 + (\varepsilon_i^2 - \varepsilon_j^2)^2 + \ldots \right]
\end{equation}
is the \bfa{squared} norm of the gradient of the abundance distributions \citep{Piskunov2002p736}, and 
\begin{equation}
  \label{eq:mdi:regul:rf}
  \mathcal{R}_f = \sum_\ell^{\ell_\mathrm{max}} \sum_m \left( \alpha_{\ell,m}^2 + \beta_{\ell,m}^2 + \gamma_{\ell,m}^2 \right) \ell^2
\end{equation}
is the regularization \bfa{functional} for the magnetic field. Its role is to prevent the code from introducing high-order modes not justified by the observational data. The coefficients $\Lambda_a$ and $\Lambda_f$ determine the contributions of $\mathcal{R}_a$ and $\mathcal{R}_f$ into the total discrepancy function $\Psi$. Their values usually are chosen on the basis of a balance between the goodness of the fit and the smoothness of the solution.

Our experiments showed that the value of the appropriate regularization parameter can be found robustly with respect to the the value of the other regularization parameter, i.e., knowing approximately $\Lambda_f$ we can accurately determine $\Lambda_a$ and vice versa. This claim has simple explanation --- the abundance distribution strongly affects the Stokes~$I$ profiles and depends only slightly in the case of moderately strong magnetic fields on the magnetic field maps, while the Stokes~$QUV$ profiles show strong dependence on the magnetic field distribution. This translates to the following algorithm. We first calculate the MDI solution on a wide grid of $\Lambda_a$ and $\Lambda_f$ (in our case the grid is $16\times 16$). To determine $\Lambda_a$ we consider how $\mathcal{D}_I$ changes with respect to $\mathcal{R}_a$ for a fixed value of $\Lambda_f$. After identifying the optimal value of $\Lambda_a$ we execute the same procedure and assess how $\mathcal{D}_Q + \mathcal{D}_U + \mathcal{D}_V$ changes with respect to $\mathcal{R}_f$ for the previously determined optimal value of $\Lambda_a$, which gives us the optimal value for $\Lambda_f$. We repeat the procedure once or twice each time with the newly found optimal values of the regularization parameters to make sure that our results are consistent. As a rule, we get consistent results after the first iteration. 

We experimentally determine an \textit{optimal} value of each regularization parameter by assessing how the corresponding discrepancy function and regularization functional change as we decrease it from some starting value that produces very smooth solution (abundance maps in the case of $\Lambda_a$, and magnetic field maps for $\Lambda_f$). An optimal value has been found when further decrease of the regularization parameter leads only to relatively small improvement in the discrepancy function and large increase \bfa{in the value of the regularization functional} ($\mathcal{R}_a$ or $\mathcal{R}_f$).

\subsection{Atmosphere model}
\label{sec:mdi:atm}
\citet{Shulyak09p879} constructed a self-consistent model atmosphere for HD\,24712, taking the stratification of chemical elements into account. The authors derived the model atmosphere from iterative fitting of the observed high-resolution spectra and spectral energy distribution data using atmospheric models that account for the effects of individual and stratified abundances. Significant part of their paper is devoted to the NLTE treatment of the formation of \ion{Pr}{ii/iii} and \ion{Nd}{ii/iii} lines, which appears to seriously influence the atmosphere structure by creating inverse temperature gradient caused by the overabundance of REEs in the upper atmospheric layers. 

The detailed, self-consistent analysis of HD\,24712 performed by \citet{Shulyak09p879} is a good basis for our MDI study. Consequently, we adopted one of their models with stratified abundances, namely model number four (see Table~3 in the aforementioned paper), which was calculated for $T_\mathrm{eff} = 7250$\,K, $\log g = 4.1$, and a scaled REE opacity for \ion{Pr}{ii} and \nd{ii} ions. We find that this model is sufficient for our work because it accurately describes the mean atmosphere and effects of stratification and does not differ significantly from the model that has scaled REE opacity for \ion{Pr}{iii} and \ion{Nd}{iii} in addition to the one for \ion{Pr}{ii} and \ion{Nd}{ii} ions. The initial abundance values of the elements that we mapped were set to the values measured from the mean spectrum; for iron we used -5.5 (in $\log(N_X/N_\mathrm{tot})$ units), for neodymium -9.0, and -7.0 for sodium.

We address one final concern that might arise from our use of single mean model atmosphere for the MDI analysis of HD\,24712. \citet{Kochukhov2012p3004} investigated this problem in the case of the magnetic Ap star $\alpha^2$\,CVn. Their experiments showed that the use of a grid of model atmospheres that reflect the local surface abundance distribution of the mapped chemical elements only slightly improves the fit to Stokes~$I$, and practically has no effect on the polarization profiles as well as on the resulting magnetic field and abundance maps. In addition to that, the abundance contrast of the most important elements Fe, Cr, and Si for HD\,24712 is much less extreme compared to $\alpha^2$\,CVn. Therefore, we proceed with using a single mean atmosphere model in our study.

\subsection{Spectral lines selection}
\label{sec:mdi:lines}

\begin{table}
  \caption{Atomic data of spectral lines used for the MDI inversion.}
  \centering
  \begin{tabular}{lccc}
  \hline
  \hline
  Ion & $\lambda (\mathrm{\AA})$ & $E_\mathrm{lo} (\mathrm{eV})$ & $\log gf$ \\
  \hline
  \ion{Fe}{i} & 4267.826 & 3.111 & $-1.174^a$ \\
  \ion{Fe}{i} & 4938.814 & 2.875 & $-1.077$ \\
  \ion{Fe}{i} & 5198.711 & 2.223 & $-2.056$ \\
  \ion{Fe}{i} & 5217.389 & 3.211 & $-1.070$ \\
  \ion{Fe}{i} & 5434.524 & 1.011 & $-2.122$ \\
  \ion{Fe}{i} & 5445.042 & 4.386 & $-0.020^a$ \\
  \ion{Fe}{i} & 6008.556 & 3.884 & $-0.986$ \\
  \ion{Fe}{i} & 6024.058 & 4.548 & $-0.120^a$ \\
  \ion{Fe}{i} & 6137.692 & 2.588 & $-1.403$ \\
  \ion{Fe}{i} & 6336.824 & 3.686 & $-0.856$ \\
  \ion{Fe}{i} & 6393.600 & 2.433 & $-1.432$ \\
  \ion{Fe}{i} & 6419.949 & 4.733 & $-0.240$ \\
  \ion{Fe}{ii} & 6432.680 & 2.891 & $-3.687$ \\
  \hline
  \ion{Nd}{iii} & 5677.179 & 0.631 & $-1.450^b$ \\
  \ion{Nd}{iii} & 5802.532 & 0.296 & $-1.710^b$ \\
  \ion{Nd}{iii} & 5851.542 & 0.461 & $-1.516^b$ \\
  \hline
  \ion{Na}{i} & 5889.951 & 0.000 & $0.108$ \\
  \hline
  \end{tabular}
  \label{tbl:mdi:lines}
  \tablefoot{The columns give the ion, central wavelength $\lambda$, excitation potential of the lower atomic level $E_\mathrm{lo}$, and oscillator strength $\log gf$.\\ ${}^{(a)}$~$\log gf$ value adjusted automatically in the inversion procedure. ${}^{(b)}$~$\log gf$ corrections added for NLTE effects.}
\end{table}

From the available spectropolarimetric observations we selected 13 \ion{Fe}{i/ii}, 3 \ion{Nd}{iii} lines and one \ion{Na}{i} line. We performed the selection on the basis of the presence of strong polarization signals in the Stokes~$QUV$ profiles, and as well as having visible variations with phase in the Stokes~$I$ profiles due to abundance inhomogeneities and the Zeeman effect. Spectral lines suffering from significant blending were not included in the line list.

The reason for using these chemical elements is due to the fact that iron and neodymium have different vertical stratification profiles and the respective Stokes profiles of their spectral lines exhibit different rotational modulation with phase. Moreover, the phase variation of the sodium lines differs from that of the iron and neodymium lines. Therefore, by using these elements we can fully probe the surface of HD\,24712 and obtain a robust reconstruction of the magnetic field and abundance distribution. Detailed analysis of the behavior of the polarization signatures of individual lines in the spectrum of HD\,24712 was presented in Sect.~4 of paper~I.

The complete line list adopted for the MDI inversion of HD\,24712 is presented in Table~\ref{tbl:mdi:lines}. The atomic data for the lines were extracted from the VALD database \citep{Kupka1999p119}. The \ion{Fe}{i} 5434.524\,\AA{} line, which is insensitive to the magnetic field was also included in the line list. This line serves as a reference that can tell us how well the abundance distribution of iron reproduces the Stokes~$I$ profiles of non-magnetic lines. During the inversion procedure it was noticed that three iron lines had incorrect $\log gf$ values. The corrections for these three lines were calculated automatically by the \invers{10} code. The three \ion{Nd}{iii} lines were chosen from an initial list of known \ion{Nd}{ii/iii} lines on the basis of our criteria. To the oscillator strengths of these lines we added NLTE corrections that were calculated by \citet{Mashonkina2005p309}.

\subsection{Optimization of $v_e \sin i$, $\Theta$ and $i$}
\label{sec:mdi:opt}
\begin{figure}
  \centering
  \includegraphics[width=0.45\textwidth]{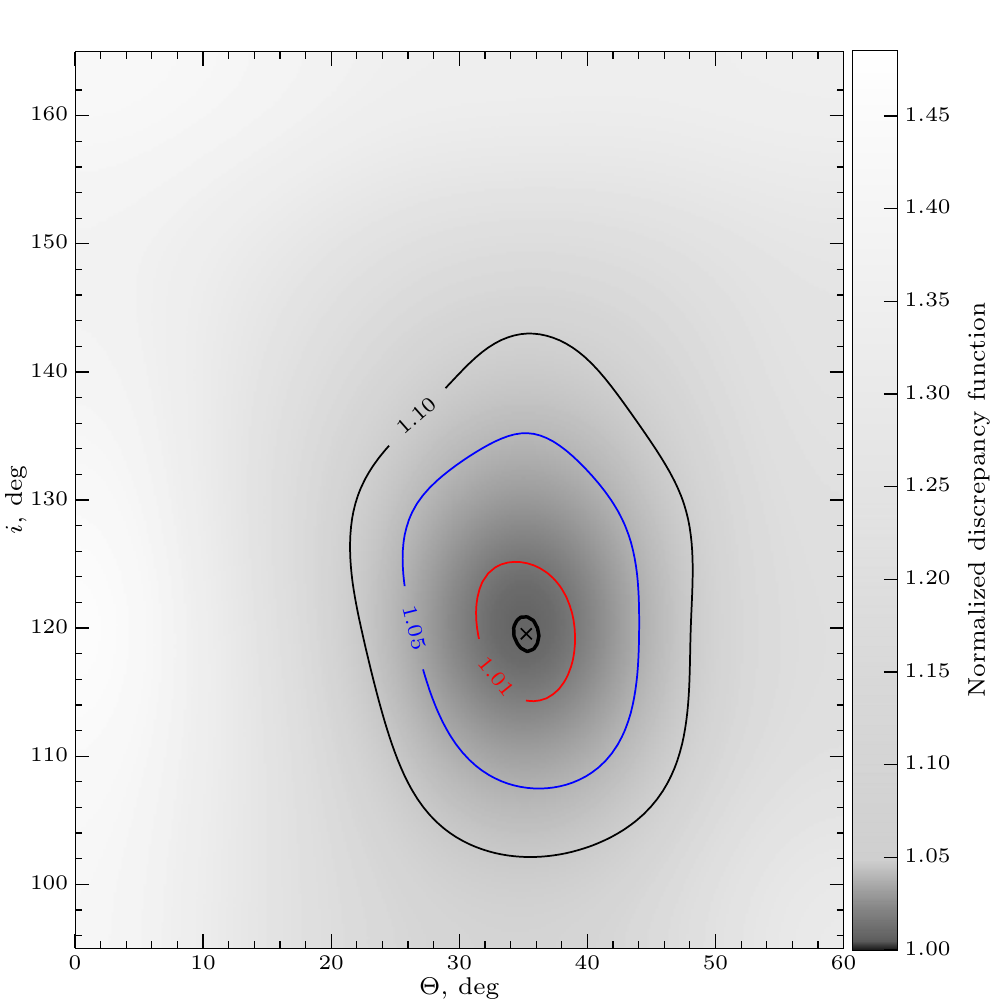}
  \caption{Variations of the normalized discrepancy function for the Stokes~$QU$ profiles, $\mathcal{D}_Q + \mathcal{D}_U$, as a function of inclination angle $i$ and azimuth angle $\Theta$. The innermost unmarked contour plotted with thick black line corresponds to the 5-sigma confidence level of the fit. The other contours marked with 1.01, 1.05, and 1.10 correspond to 1\%, 5\%, and 10\% increase of the discrepancy function from its minimal value.}
  \label{fig:mdi:angles}
\end{figure}

\begin{figure*}
  \centering
  \includegraphics[width=0.4975\textwidth]{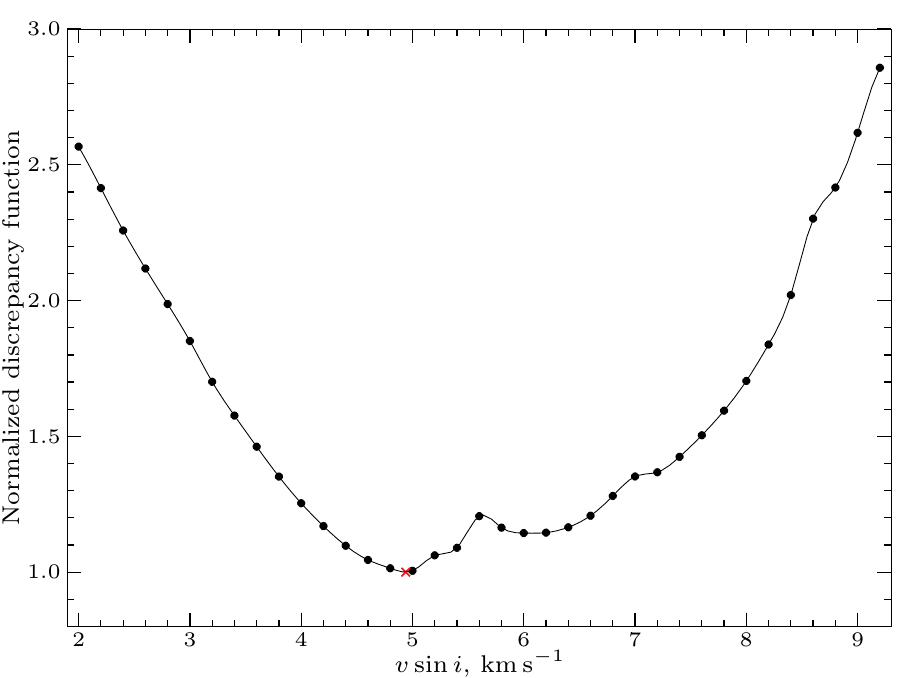}
  \includegraphics[width=0.4975\textwidth]{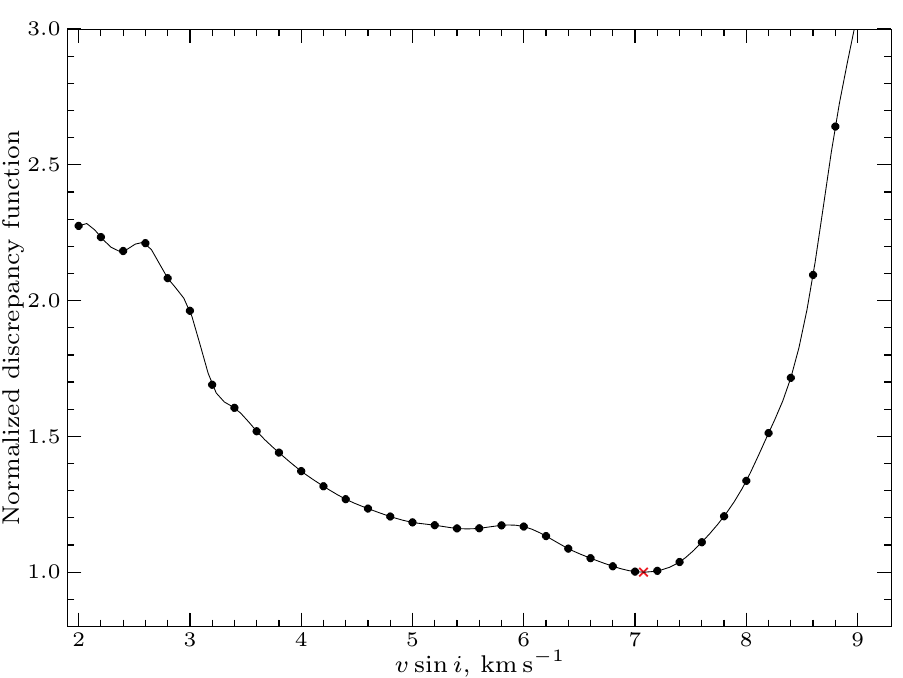}
  \caption{The normalized discrepancy function $\mathcal{D}_I$ for the iron lines (left panel) and for the neodymium lines (right panel) as a function of projected equatorial velocity $v_e \sin i$ (symbols). The solid curve is an interpolating cubic spline used to find the minimum of the discrepancy functions. The ``x'' symbol on the plot marks the position of the minimum.}
  \label{fig:mdi:vsini}
\end{figure*}

Magnetic Doppler imaging in four Stokes parameters in addition to the stellar atmosphere model and the line list requires the orientation of the stellar rotation axis and the projected rotational velocity of the star. The orientation of the stellar rotation axis is determined by the inclination~$i$ and azimuth angle $\Theta$. The inclination is defined as the angle between the rotation axis and the line-of-sight to the observer; it changes from $0^\circ$ to $180^\circ$. The azimuth angle can have values in the range $[0^\circ,360^\circ]$, and it determines the position angle of the sky-projected rotation axis. Note that because the Stokes~$QU$ profiles are changing with $2\Theta$ there is an ambiguity between $\Theta$ and $\Theta + 180^\circ$.

The initial values $i=138^\circ$ and $\Theta=29^\circ$ used here were determined in paper~I, where we combined our longitudinal field and net linear polarization measurements with available broadband linear polarization data \citep{Leroy1995p79} and computed the parameters of the so-called canonical model \citep{Landolfi1993p285} for dipolar magnetic field geometry. We adopted 5.6\,\kms{} \citep{Ryabchikova97p1137} as initial value for the projected rotational velocity. 

Using numerical experiments, \citet{Kochukhov2002p868} showed that incorrect values of these parameters lead to higher values of the discrepancy function. The optimal value of the rotational velocity can be determined by comparing the discrepancy function for the Stokes~$I$ profiles calculated with various values of $v_e \sin i$. Incorrect values of $i$ and $\Theta$ lead to similar increase of the discrepancy function for Stokes~$Q$ and $U$ profiles.

For the optimization of $i$ and $\Theta$ we computed 42 MDI solutions on a grid $i \in [100^\circ, 160^\circ]$ and $\Theta \in [5^\circ, 55^\circ]$ with $10^\circ$ steps for both angles. The resulting normalized discrepancy function as a function of $i$ and $\Theta$ is plotted in Fig.~\ref{fig:mdi:angles}. The minimum value of the discrepancy function was found for $i=120^\circ$ and $\Theta = 35^\circ$. We compared the Stokes~$QU$ profiles for the optimal values of $i$ and $\Theta$ to the ones computed for values greater or smaller by $10^\circ$, which is the grid step in our case. The comparison showed that with the new values of $i$ and $\Theta$ our MDI procedure appears to better reproduce the observed Stokes~$QU$ profiles. We accept the newly found values of $i$ and $\Theta$ as final for the rest of the paper.

The rotational velocity was determined in a similar matter. We produced 37 individual MDI solutions for $v_e \sin i$ in the range from 2\,\kms{} to 9\,\kms{}. We computed the discrepancy function for the Stokes~$I$ profiles separately for the Fe, and for the Nd lines. This is necessary because Nd lines require additional broadening for the Stokes~$I$ profiles, which is not present in the Fe lines. This phenomenon is known for other Ap stars \citep[see][]{Ryabchikova2007p907}. In Fig.~\ref{fig:mdi:vsini} we illustrate the normalized discrepancy function for Stokes~$I$ profiles of the Fe and Nd lines accordingly. From this figure it can be seen that the discrepancy function for the Fe lines shows complex behavior with two close minima in the 5 -- 6\,\kms{} range with a peak around 5.5\,\kms{}. This behavior of the discrepancy function for the Fe lines leads us to believe that we cannot with confidence find optimized $v_e \sin i$. We also draw your attention to the significantly higher $v_e \sin i \simeq 7.1$\,\kms{} derived from Nd lines. Despite being less sensitive to changes in $v_e \sin i$ one can use the discrepancy functions for the rest of the Stokes parameters in a similar manner to get an understanding of the range where $v_e \sin i$ is the most probable. The discrepancy functions for these Stokes parameters showed well defined minima in the range 5 -- 6\,\kms{}. As a result, we decided to adopt new value for $v_e \sin i = 5.5\pm0.5$\,\kms{} corresponding to the center of that range. In the rest of the paper this is value of $v_e \sin i$ that is used for all other MDI inversions.

\section{Results}
\label{sec:res}

\subsection{Individual vs. simultaneous MDI of chemical elements}
\begin{figure*}
  \centering
  \includegraphics[width=0.85\textwidth]{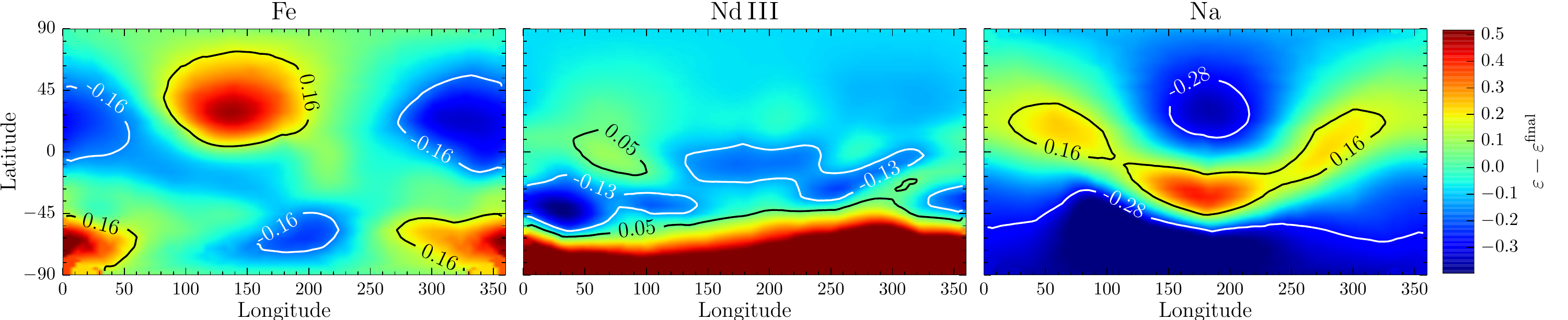}
  \caption{Differences in the abundance distribution of Fe, \nd{iii}, and Na from their \textit{individual} mapping relative to the corresponding maps from the \textit{simultaneous} MDI of these elements. We note that for Na mapping we used the magnetic field maps derived from the simultaneous MDI inversion of the Fe and \nd{iii} lines. On each panel with \textit{white} and \textit{black} lines we  marked the 16 and 84 percentiles. The bar on the right indicates the abundance difference in $\log (N_X / N_\mathrm{tot})$ units of element $X$.}
  \label{fig:diffs_abn}
\end{figure*}

\begin{figure*}
  \centering
  \includegraphics[width=0.85\textwidth]{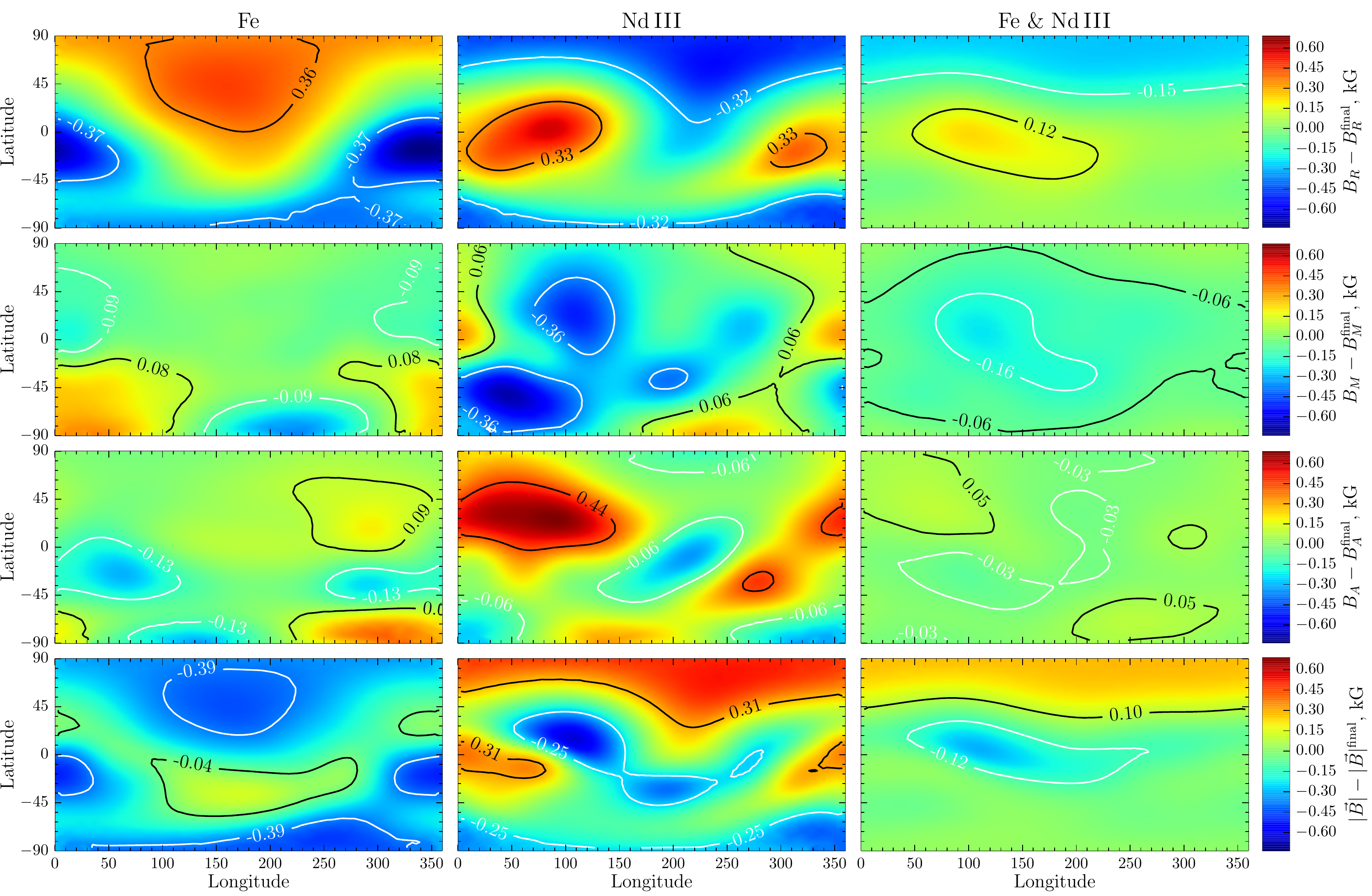}
  \caption{Differences in distribution of the magnetic field components relative to the corresponding maps from the \textit{simultaneous} MDI of Fe, \nd{iii} and Na. In rows one to four we plotted the differences for the radial, meridional, and azimuthal components of the magnetic field, and for the field modulus. In column one we plotted the differences for the magnetic field components derived from the MDI inversion of Fe; in column two we plotted the same quantities for \nd{iii}; in column three we plot the differences of the magnetic field components derived from the simultaneous MDI of Fe and \nd{iii} lines. On each plot with \textit{white} and \textit{black} lines we marked the 16 and 84 percentiles. The bars on the right indicate the difference of the corresponding components measured in kG.}
  \label{fig:diffs_mf}
\end{figure*}

Our magnetic Doppler imaging study was separated into three steps. The \textbf{first step} involved producing maps of the magnetic field and chemical abundance from \textit{individual} mapping of Fe and \nd{iii}. In the \textbf{second step} we produced magnetic field and abundance distribution maps from \textit{simultaneous} mapping of Fe and \nd{iii}. Additionally, we used the magnetic field maps produced in this step and calculated abundance map for Na. In the \textbf{third step} we performed \textit{simultaneous} MDI mapping of the three chemical elements Fe, \nd{iii}, and Na. The magnetic field and abundance distribution maps calculated in the third step are called ``the final maps'' in the rest of the paper.

In this section we investigate how robust are results produced from \textit{individual} versus \textit{simultaneous} MDI of Fe, \nd{iii}, and Na. Ideally, all maps, whether they are produced from individual or simultaneous mapping of chemical elements, should show same results. However, due to inhomogeneous surface distribution of chemical elements in the atmospheres of Ap stars spectral lines of different elements can be sensitive to certain parts of the surface of the star by a different amount. Therefore, it is necessary to have a line list that probes the entire stellar surface and not just parts of it. Our line list, discussed in Sect.~\ref{sec:mdi:lines} is the result of this argument. We note that while the MDI analysis for Fe and \nd{iii} in steps one and two was self-consistent, i.e., we derived abundance and magnetic field maps, for the single Na line we used the already derived magnetic field maps from simultaneous MDI of Fe and \nd{iii}. That way we can test how the single Na line affects our final results. The reasoning behind this is that mapping the magnetic field and abundance distribution from one single line can give biased results. Additionally, our Na line cannot be fitted satisfactorily by the \invers{10} code, which might raise questions about the quality and robustness of our inversions. 

We compare two maps $M_1$ and $M_2$ calculated on the same surface grid by calculating the difference, $M_1 - M_2$, from which we compute the median value, and 16 and 84 percentiles that in the case of normal distribution are the same as one standard deviation below and above the mean value. We prefer using the 16 and 84 percentiles and the median instead of the classical statistical estimators --- the mean and standard deviation --- with the goal of minimizing possible effects of long non-normal tails sometimes present in MDI maps.

We also performed a visual comparison of the morphological features of the two maps. In the case of magnetic field maps we also assess the \textit{relative energy} of their poloidal and toroidal components as a function of spherical harmonics number $\ell$ (see Sect.~\ref{sec:mdi:intro}). We do this as a way to compare the magnetic field geometry of different solutions. 

\begin{figure}
  \centering
  \includegraphics{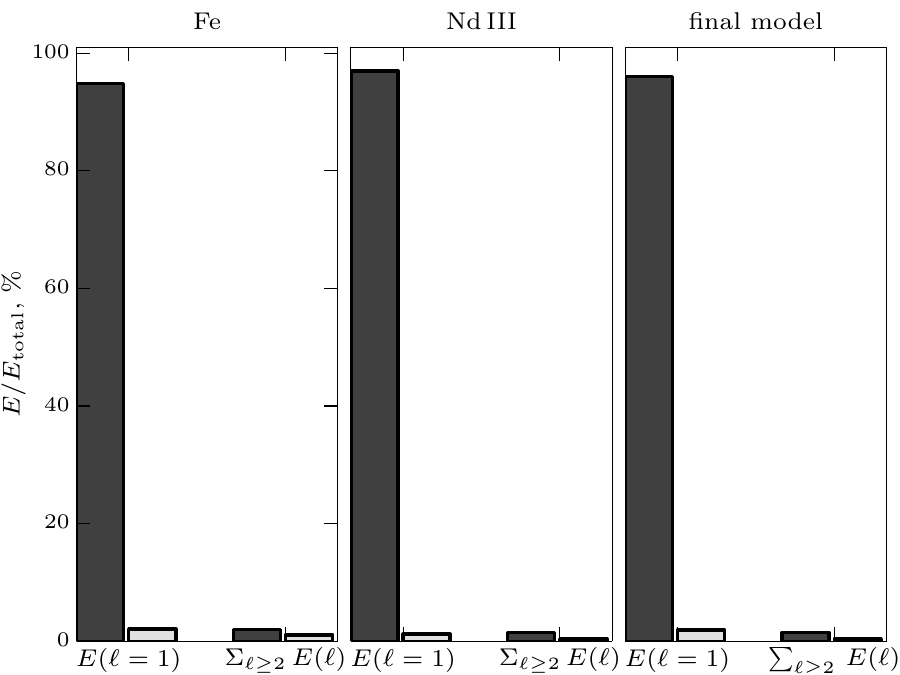}
  \caption{Relative energies of the poloidal and toroidal harmonic modes for the magnetic field topology of HD\,24712. In panels one to three we show relative energies of the magnetic field from individual MDI of Fe, \nd{iii}, and from simultaneous MDI of Fe, \nd{iii}, and Na (``final model''). The energy of the poloidal and toroidal modes in each panel are shown in dark and light gray. The first bar in each panel for the corresponding energy mode (poloidal, toroidal) is for $\ell=1$; the second such bar represents the sum for all energies with $\ell \geq 2$.}
  \label{fig:energy}
\end{figure}

The differences of abundance distribution maps of Fe, \nd{iii}, and Na relative to the corresponding final maps can be found in Fig.~\ref{fig:diffs_abn}. The analogous quantities for the magnetic field maps derived from individual mapping of Fe and \nd{iii} relative to the respective final magnetic field maps are presented in Fig.~\ref{fig:diffs_mf}. Fig.~\ref{fig:energy} compares the energies of the poloidal and toroidal harmonic modes.

Let us consider the resulting difference maps for the abundance distribution of the mapped chemical elements. The analysis of the differences in the distribution of Fe derived from individual mapping relative to its final map shows good agreement between the two. The difference map has median value of $-0.02$\,dex and shows no statistically significant systematic shifts relative to the final map. Furthermore, comparison of the morphological properties of the two Fe maps shows no significant differences --- both recovered maps have similar features. 

The \nd{iii} difference map shows similar agreement as the Fe maps, with median value of $-0.07$\,dex. There is one almost circular area around the visible magnetic pole with approximately 0.5\,dex overabundance relative to the final \nd{iii} map that deserves closer consideration. Taking into account its relatively small surface area and the large effective abundance range of the final map of 2.6\,dex we suggest that this spot has small influence on the final results.

Finally, the abundance map produced for Na shows no systematic shifts from the final Na map with median value of $-0.05$\,dex. Although, in this case, the difference map shows larger deviations of the individual map from the final one by about 0.3\,dex, this is still much smaller than the effective abundance range of 3.5\,dex for the final map. Thus, we consider the Na map to be reproduced robustly.

The difference maps for the magnetic field provide a more complicated story. For Fe we can see that the recovered magnetic field tends to be weaker by about 0.22\,kG as given by the median value of the difference for the field modulus. This is especially noticeable in Fig.~\ref{fig:diffs_mf} (row 4, column 1). Despite the weaker field strength, 94.5\% of the energy of the field is located in the first harmonic of the poloidal field component (Fig.~\ref{fig:energy}, panel~1). This indicates that by using only Fe lines for the derivation of the magnetic field maps of HD\,24712 we risk getting a weaker field.
 
The difference maps of the magnetic field derived from \nd{iii} lines show variation by about 0.35\,kG for all components of the field. Despite this, the energy of the magnetic field derived from the individual mapping of \nd{iii} is almost identical to the one from Fe. The strongest component is the poloidal one for the first harmonic with 97\% of the total magnetic field energy (Fig.~\ref{fig:energy}, panel~2) versus 96\% for the same component in the final map.

Finally, the magnetic field maps derived from simultaneous MDI of Fe and \nd{iii} show very small ($< 0.15$\,kG) deviations from the appropriate final maps. Therefore, adding the single Na line to our line list does not significantly change our final maps. 

In summary, we can make the following conclusions:
\begin{enumerate}
\item The abundance maps of Fe, Nd, and Na computed from their \textit{individual} inversions have the same morphology and show no significant systematic shifts relative to the corresponding final maps computed from \textit{simultaneous} MDI inversion. The discrepancy between the respective maps is around 0.15\,dex for Fe and Nd, and less than 0.3\,dex for Na.
\item We constrain different components of the magnetic field of HD\,24712 with an accuracy of approximately 0.3\,kG. 
\item The use of single elements in MDI inversions can lead to a situation where the reconstructed magnetic field has systematic bias. Our analysis on the example of HD\,24712 emphasizes that a proper MDI study of Ap stars should be done with a set of lines that probes the entire surface of the star.
\item The energy of the magnetic field as a function of the spherical harmonic degree $\ell$ changes by around 1\%, whether derived from \textit{individual} or \textit{simultaneous} MDI inversion of Fe and Nd, which implies that the global magnetic field geometry of HD\,24712 is constrained robustly in both cases.
\end{enumerate}

\subsection{Maps from simultaneous MDI of Fe, \nd{iii}, and Na}
\begin{figure*}
  \centering
  \includegraphics[page=1,width=0.75\textwidth]{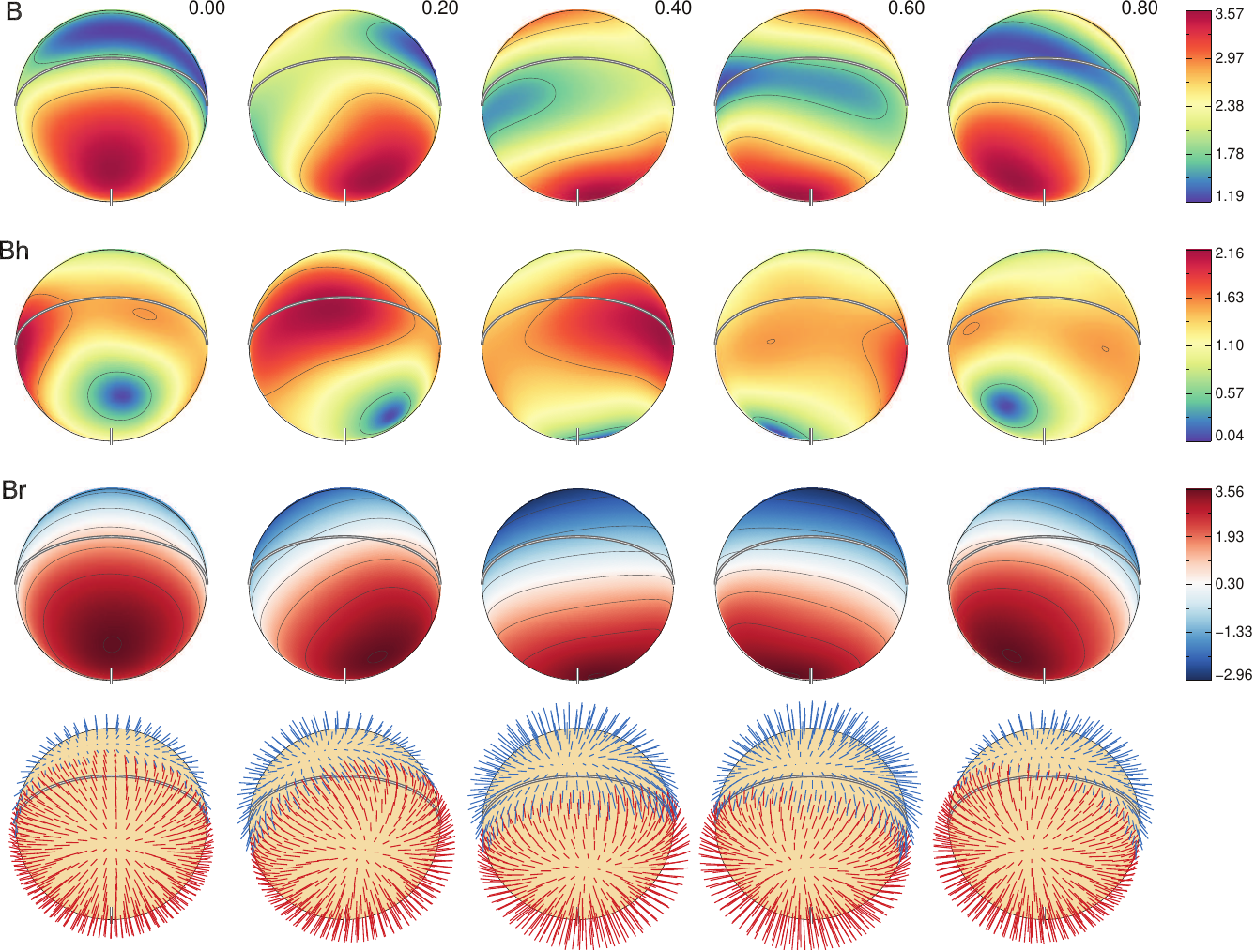}
  \caption{Distribution of the magnetic field on the surface of HD\,24712 derived from simultaneous MDI analysis of Fe, \nd{iii}, and Na. The plots show the distribution of magnetic field modulus (\textit{first row}), horizontal field (\textit{second row}), radial field (\textit{third row}), and field orientation (\textit{fourth row}) on the surface of HD\,24712. The bars on the right indicate field strength in kG. The contours are plotted with 1\,kG step. The arrow length in the bottom plot is proportional to the field strength. The star is shown at five rotational phases, indicated above the spherical plots.}
  \label{fig:mf}
\end{figure*}

\begin{figure*}
  \centering
  \includegraphics[page=2,width=0.75\textwidth]{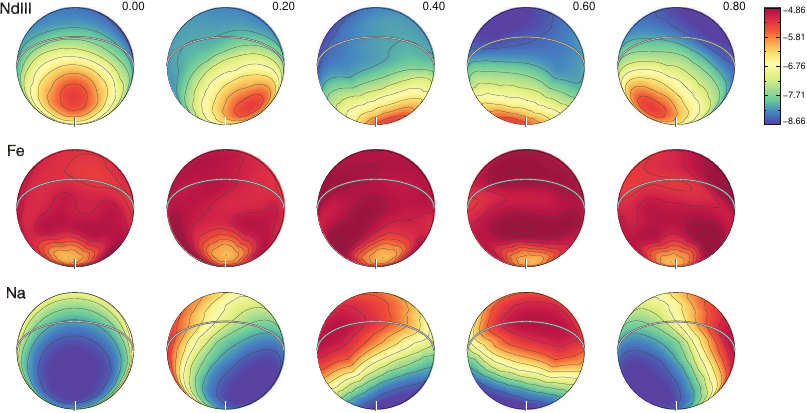}
  \caption{Abundance distribution of \nd{iii}, Fe, and Na on the surface of HD\,24712. The \textit{first} row shows the surface map of the \nd{iii}, the \textit{second} and \textit{third} rows illustrate the surface abundance maps of Fe and Na respectively. These maps were derived from the \textit{simultaneous} mapping of the three elements. The bars on the right next to each panel indicate the abundance in $\log (N_X / N_\mathrm{tot})$ units of element $X$. The contours for are plotted with a step of 0.4\,dex \nd{iii} and Na, and 0.2\,dex for Fe. The vertical bar indicates the rotation axis.}
  \label{fig:abn}
\end{figure*}

\begin{figure*}
  \centering
  \includegraphics[width=\textwidth,page=1]{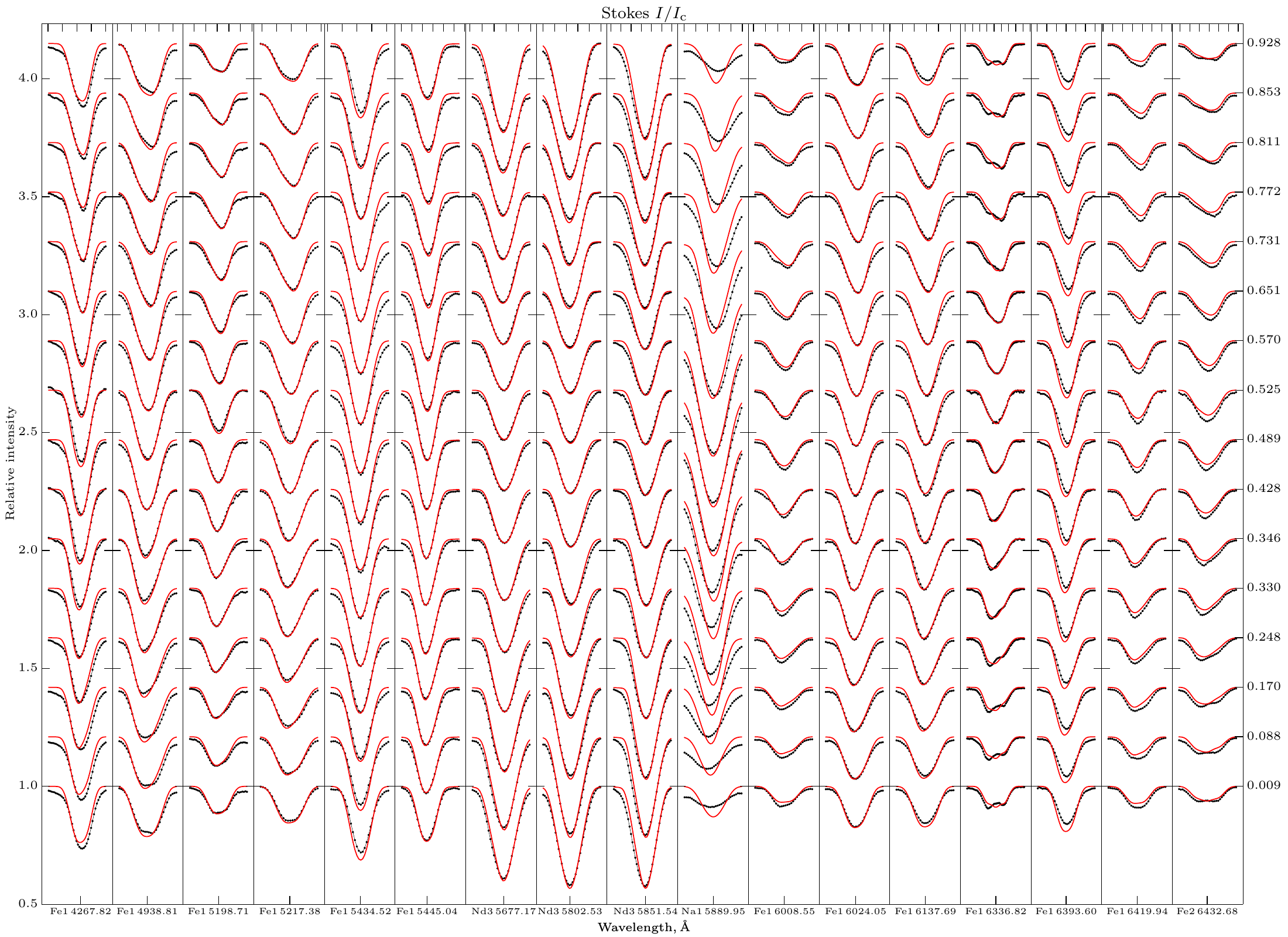}
  \caption{Comparison of the observed (\textit{dots connected with lines}) and synthetic (\textit{lines}) Stokes~$I$ profiles calculated for the final magnetic field and abundance maps (Fig.~\ref{fig:abn} and~\ref{fig:mf}) for all lines used in the MDI inversion. The distance between two consecutive ticks on the horizontal top axis of each panel is 0.1\,\AA{} indicating the wavelength scale. Rotational phases are indicated in the right of the figure.}
  \label{fig:prfs}
  \label{fig:prfs_1}
\end{figure*}

\begin{figure*}
  \centering
  \includegraphics[width=\textwidth,page=2]{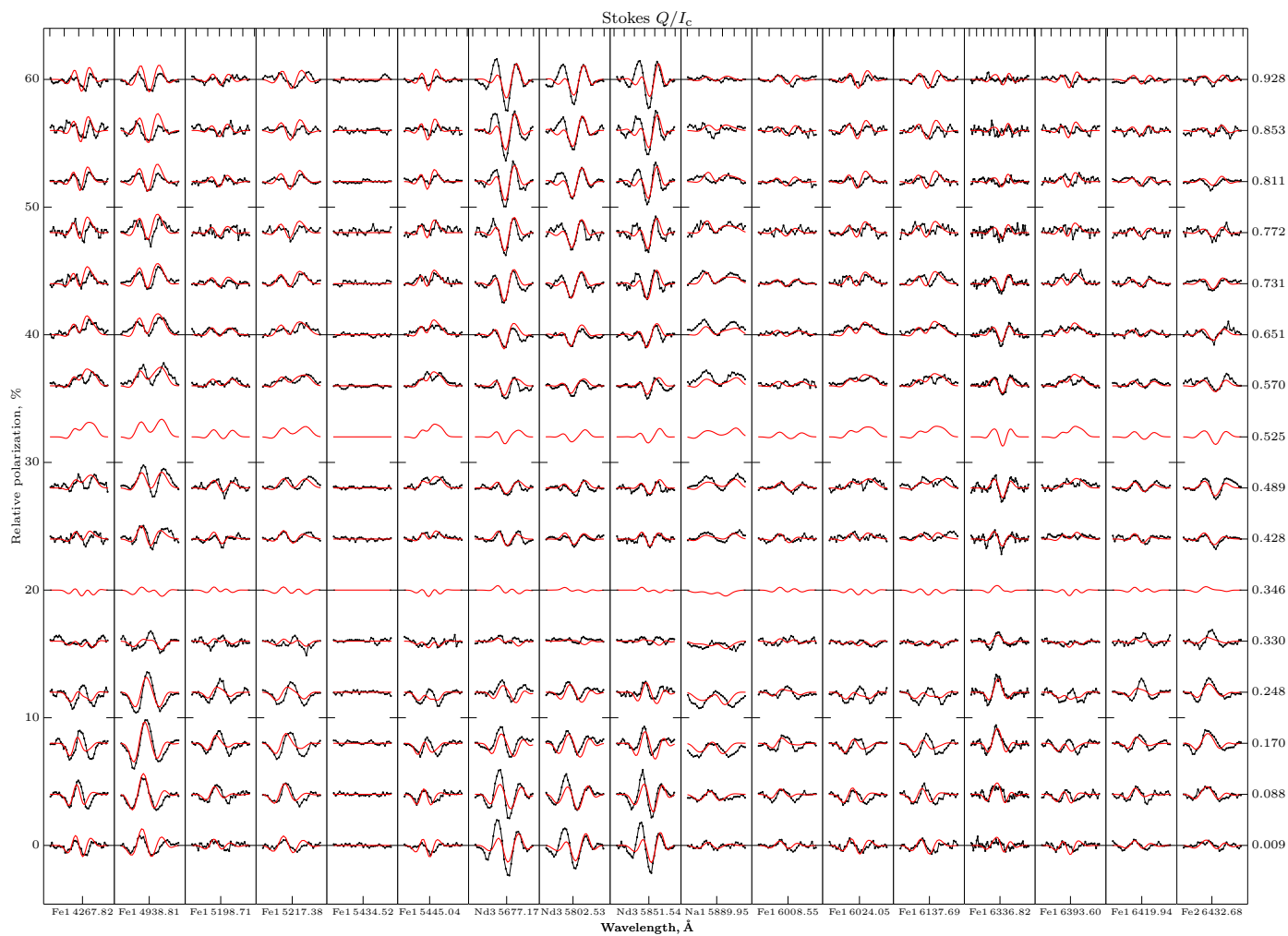}
  \caption{Same as Fig.~\ref{fig:prfs_1} for the Stokes~$Q$ profiles.}
  \label{fig:prfs_2}
\end{figure*}

\begin{figure*}
  \centering
  \includegraphics[width=\textwidth,page=3]{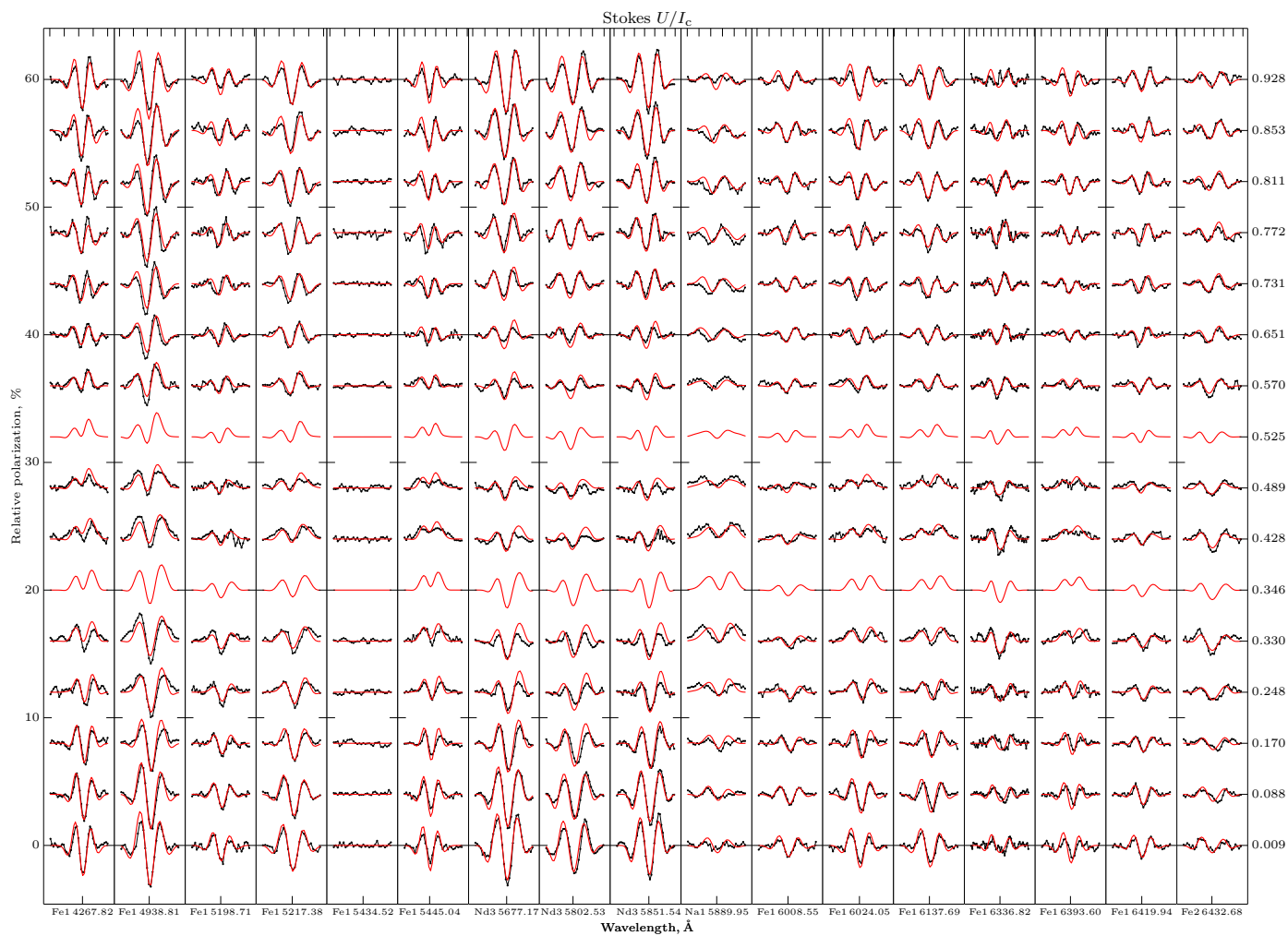}
  \caption{Same as Fig.~\ref{fig:prfs_1} for the Stokes~$U$ profiles.}
  \label{fig:prfs_3}
\end{figure*}

\begin{figure*}
  \centering
  \includegraphics[width=\textwidth,page=4]{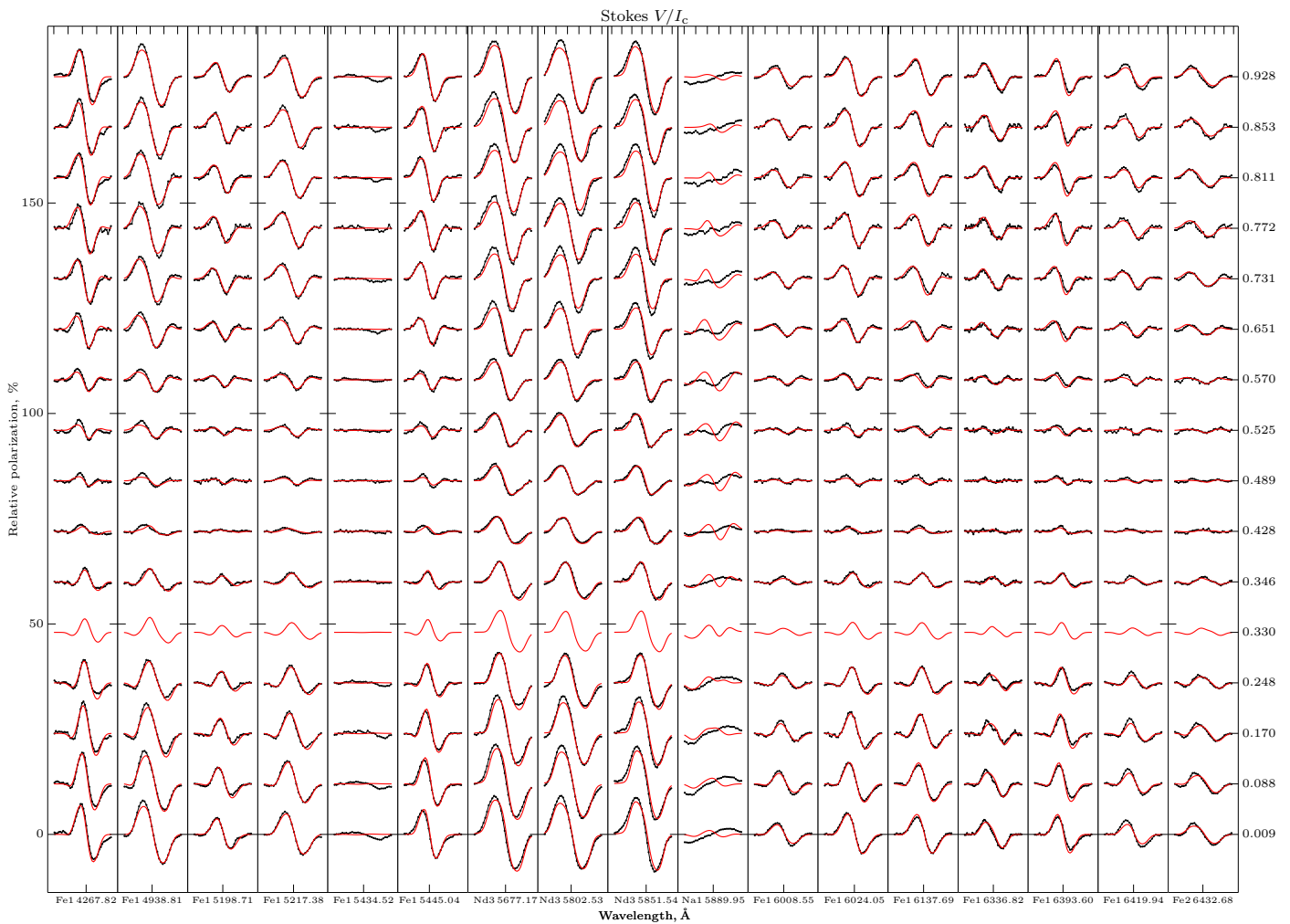}
  \caption{Same as Fig.~\ref{fig:prfs_1} for the Stokes~$V$ profiles.}
  \label{fig:prfs_4}
\end{figure*}

The final magnetic field and abundance maps derived from the simultaneous mapping of Fe, \nd{iii}, and Na are shown in Fig.~\ref{fig:mf} and~\ref{fig:abn}. In Fig.~\ref{fig:mf} we plot the spherical projection of the field modulus $(\sqrt{B_r^2 + B_m^2 + B_a^2})$, the strength of the horizontal $(\sqrt{B_a^2 + B_m^2})$ and radial field $B_r$ components; the bottom row shows the vector magnetic field. The comparison of the observed and the calculated line profiles for the entire line selection is presented in Figs.~\ref{fig:prfs_1}--\ref{fig:prfs_4}. 

The reconstructed magnetic field of HD\,24712 appears to be mostly poloidal and dipole-like. The poloidal harmonic component with $\ell=1$ dominates over all other toroidal and poloidal harmonics, containing 96\,\% of the total magnetic field energy. The contribution of all other modes with $\ell \geqslant 2$ is much less significant as shown in Fig.~\ref{fig:energy} (panel~3).

We report that our final magnetic field maps show slight \bfa{asymmetry} between the positive (visible) and negative magnetic pole, see Fig.~\ref{fig:mf}, of approximately 0.6\,kG. We believe that this discrepancy does not represent sufficient evidence that the magnetic field topology of HD\,24712 has strong deviations from axial symmetry. It is possible that this is caused purely by the nature of the magnetic field configuration of the star, where only the positive magnetic pole is visible by an observer from Earth. 

We conclude that the magnetic field topology of HD\,24712 has a dominant dipolar component with a very weak contributions at smaller spatial scales from higher-order harmonics. This result contrasts the finding of roughly dipole-like global field with strong small-scale features for $\alpha^2$\,CVn \citep{Kochukhov10p13}, and the quasi-dipolar field of 53\,Cam, which has mostly dipolar poloidal component with toroidal contributions of similar strength on spatial scales of $30^\circ\text{--}\,40^\circ$ \citep{Kochukhov04p13}.

The abundance distribution of Fe and \nd{iii} for HD\,24712 was previously derived by \citet{Lueftinger10p71} from MDI study of Fe and \nd{iii} lines using only Stokes~$I$ and $V$ parameters. The new maps confirm the previous findings, and also show some details not present in the previous study. 

The abundance distribution of Fe varies between -5.53 and -4.86\,dex on the $\log (N_X/N_\mathrm{total})$ scale. One new detail is the appearance of a roughly circular area around the positive magnetic pole that is strongly depleted relative to the median value for around 0.3\,dex. The new \nd{iii} abundance map shows significantly stronger variations changing from -8.66 to -6.04\,dex. This corresponds to a range of values of 2.61\,dex versus the 1.1\,dex reported in the study by \citet{Lueftinger10p71}. The range of the abundance values for Fe is 0.67\,dex and matches the one reported in the previous study.

We also found a discrepancy in Nd spot location. \citet{Lueftinger10p71} detected a possible longitudinal offset, amounting to 0.04 of the rotation period, of the position of the maximum of the \nd{iii} abundance map relative to the maximum of the magnetic field. Our MDI study shows that the longitudinal offset for \nd{iii} is very small if present at all. We attribute this to the better quality of our observational material and the denser phase sampling resulting in a higher accuracy when determining longitudinal position of abundance spots; \citet{Lueftinger10p71} could determine longitudinal positions of abundance structures with precision of 0.02--0.05 rotational periods. We also found that our adopted period, which is slightly smaller than the one used by \citet{Lueftinger10p71}, results in a positive longitudinal shift of around 0.01 rotational periods relative to the previous study. Therefore, we believe that the absence of visible longitudinal offset of the maximum of the \nd{iii} abundance map relative to the position of the maximum of the magnetic field in this work is caused by a combination of the higher quality of the observations and revised rotational period used here.

The maps of the abundance distribution of Fe and \nd{iii} (Fig.~\ref{fig:abn}) in combination with the magnetic field maps (Fig.~\ref{fig:mf}) reproduce the Stokes profiles of the spectral lines reasonably well. However, certain discrepancies are present if we look closely at minor spectral details. This can be seen on the Stokes~$I$ line profiles of \fe{i}\,6336.824\,\AA{}. This line has a simple Zeeman splitting pattern and is highly sensitive to the magnetic field, which results in the appearance of two partially resolved components that are easily visible in the observed Stokes~$I$ profiles (Fig.~\ref{fig:prfs_1}), however the synthetic line profiles do not show such splitting. The second example that exhibits similar behavior is the \fe{ii}\,6432.68\,\AA{} line, for which the \invers{10} code also does not produce the characteristic splitting in the Stokes~$I$ profiles.

Another discrepant behavior is observed in the Stokes~$Q$ profiles of the \nd{iii} lines, for which the \invers{10} code cannot reproduce the blue wing of the profiles around maximum magnetic field. This problem persists even when we perform MDI inversion using only \nd{iii} lines, although the discrepancy is less pronounced.

It is our opinion that these discrepancies do not affect our results significantly. In the case of Fe lines with visible Zeeman splitting in the Stokes~$I$ profiles, their general behavior with phase is reproduced on a satisfactory level. The observed behavior in the calculated Stokes~$Q$ profiles of the \nd{iii} lines can be caused by the Stokes~$Q$ profiles of Fe lines, which are systematically weaker and more numerous than \nd{iii} line profiles (13 versus 3) and thus contribute more to the MDI solution.

Here, for the first time for HD\,24712, we made an attempt to derive an abundance map of Na. The resulting abundance map explains the behavior of Na lines fairly well, whose profiles change in anti-phase to Nd lines. The abundance map of this element shows strong horizontal gradient, its range of values is 3.49\,dex and changes from -8.65 to -5.16\,dex. The morphological characteristics of the Na abundance map contrasts the one for \nd{iii}; Na is overabundant near the negative magnetic pole, which is invisible to the observer from Earth, and is depleted near the positive magnetic pole. Such abundance distribution of Na explains the peculiar behavior of the circular polarization profile of the \sod{i}\,5889.951\,\AA{} line, which changes sign around zero phase. However, special care has to be taken when interpreting these results. In Figs.~\ref{fig:prfs_1}-\ref{fig:prfs_4} we see that our best fit for this line does not properly describe its behavior around zero phase corresponding to magnetic maximum. Our further analysis of this line shows that in order to fully reproduce the behavior of the line profiles of \sod{i}\,5889.951\,\AA{} with phase we need to allow the \invers{10} code to produce an abundance map with a range of values of 5.6\,dex compared to the \bfa{3.5\,dex} for the map presented in Fig.~\ref{fig:abn}. For such an abundance map the regularization parameter $\Lambda_a$ is non-optimal (see Sect.~\ref{sec:mdi:intro}).

We suspect that the reason for the insufficiently good fit of the \sod{i} line might be the presence of strong vertical stratification complicated by NLTE effects. In our current version of the \invers{10} code we do not account for these effects. We plan to investigate vertical stratification structures in a future paper.


\subsection{Dipolar field parameters}
In paper~I we considered our longitudinal magnetic field and net linear polarization measurements inferred from the LSD profiles of HD\,24712 together with available broadband linear polarization measurements obtained by \citet{Leroy1995p79} in the framework of the so-called canonical model introduced by \citet{Landolfi1993p285}, and derived the global magnetic field parameters in the case of purely dipolar magnetic field geometry. The comparison of the global field parameters with the previous result obtained by \citet{Bagnulo1995p459} showed good agreement between them with the exception of a somewhat lower $B_p$ obtained by us.

Considering that in this work we use spherical harmonics decomposition of the magnetic field, it is possible to use the coefficients of the decomposition that correspond to the poloidal field for $\ell=1$ to derive the dipole field strength $B_p$ and the angle $\beta$ between the rotation axis and the magnetic axis. Using this approach we derived $B_p=3439$\,G and $\beta=160^\circ$ from the solution that was computed from the simultaneous mapping of Fe, \nd{iii}, and Na.

The comparison of the newly derived values for the dipolar magnetic field parameters with the ones presented in paper~I shows that the new value for $B_p$ is within three sigma of the previous measurement; the slightly higher value of $\beta$ by about $15^\circ$ relative to its previous value may be related to the revised value for the inclination angle $i=120^\circ$ that was adopted in this study (see Sect.~\ref{sec:mdi:opt}). 

We emphasize that the values for $B_p$ and $\beta$ were derived using only a \bfa{subset} of the spherical harmonics coefficients derived in our study, and therefore, represent a simplified picture of the magnetic field of HD\,24712.


\subsection{Stokes profiles and surface distribution of calcium} 
\begin{figure*}
  \centering
  \includegraphics[page=3,width=0.75\textwidth]{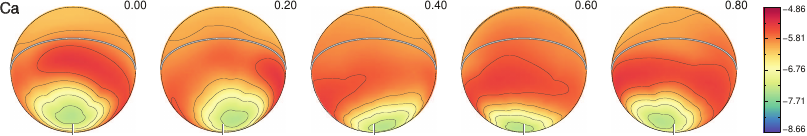}
  \caption{Abundance distribution of Ca on the surface of HD\,24712. The bar on the far right next denotes the abundance in $\log (N_{Ca} / N_\mathrm{tot})$ units. The contours are plotted with a step of 0.4\,dex. The vertical bar on each projection indicates the rotation axis. Phase is indicated in the upper right corner of each projection.}
  \label{fig:abn_ca}
\end{figure*}

\begin{figure*}
  \centering
  \includegraphics[width=0.4\textwidth, height=0.45\textheight, page=1]{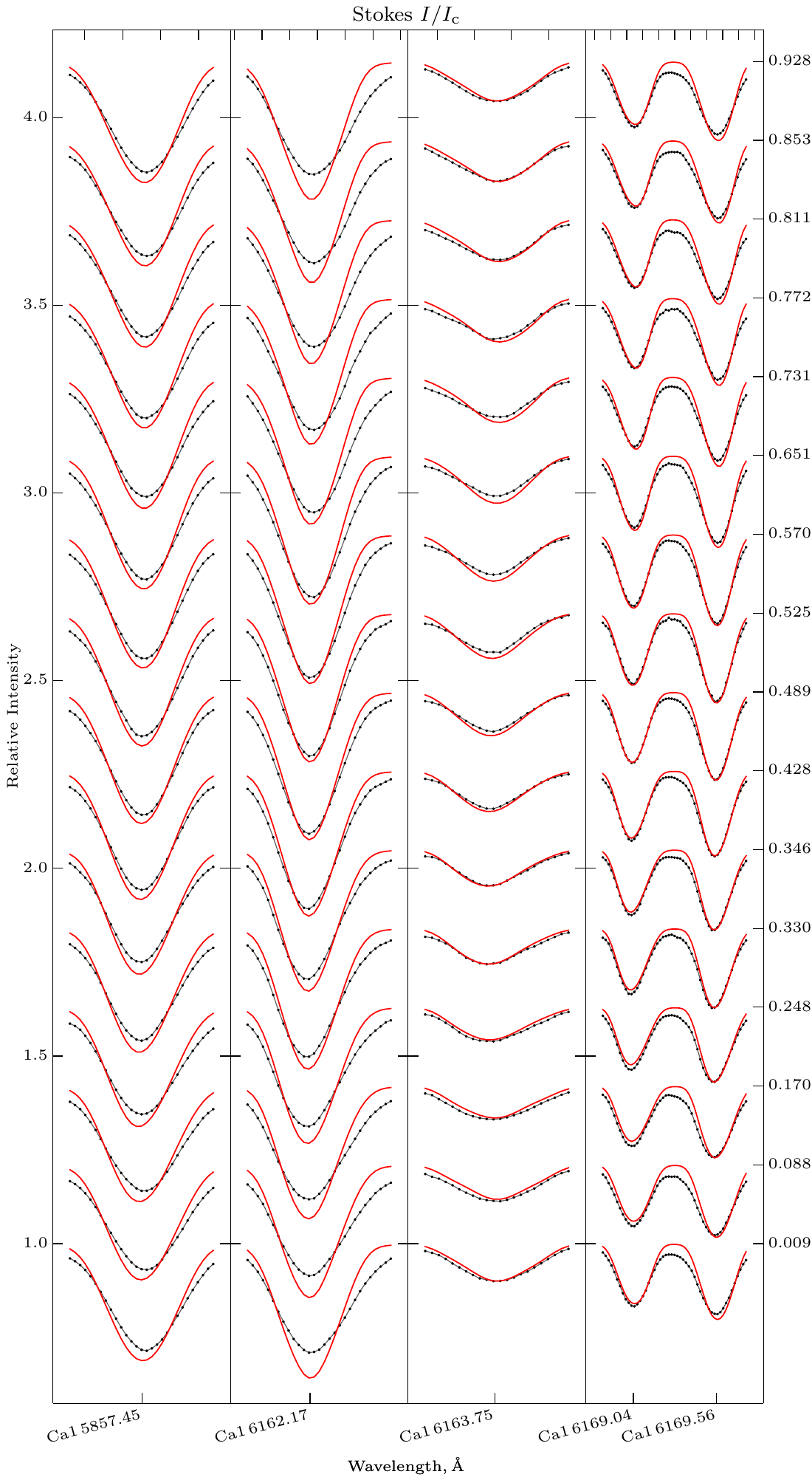}
  \includegraphics[width=0.4\textwidth, height=0.45\textheight, page=4]{Ca-lines_v3p1_207.pdf}\\
  \includegraphics[width=0.4\textwidth, height=0.45\textheight, page=2]{Ca-lines_v3p1_207.pdf}
  \includegraphics[width=0.4\textwidth, height=0.45\textheight, page=3]{Ca-lines_v3p1_207.pdf}
  \caption{Comparison of the observed (\textit{dots connected with lines}) and synthetic (\textit{lines}) Stokes profiles for Ca calculated for the final magnetic field (Fig.~\ref{fig:mf}) and abundance distribution (Fig.~\ref{fig:abn_ca}). The distance between two consecutive ticks on the horizontal top axis of each panel is 0.1\,\AA{} indicating the wavelength scale. Rotational phases are indicated on the right of each panel.}
  \label{fig:prfs_ca}
\end{figure*}

\begin{table}
  \caption{Atomic data of spectral lines used for the MDI inversion of calcium.}
  \centering
  \begin{tabular}{lccc}
  \hline
  \hline
  Ion & $\lambda (\mathrm{\AA})$ & $E_\mathrm{lo} (\mathrm{eV})$ & $\log gf$ \\
  \hline
  \ion{Ca}{i} & 5857.451 & 2.933 & $ 0.240$ \\
  \ion{Ca}{i} & 6162.173 & 1.899 & $-0.090^a$ \\
  \ion{Ca}{i} & 6163.755 & 2.521 & $-1.286^a$ \\
  \ion{Ca}{i} & 6169.042 & 2.523 & $-0.797$ \\
  \ion{Ca}{i} & 6169.563 & 2.526 & $-0.478^a$ \\
  \hline
  \end{tabular}
  \label{tbl:mdi:lines_ca}
  \tablefoot{Same as Table~\ref{tbl:mdi:lines} for the Ca line list.}
\end{table}
\citet{Lueftinger10p71} attempted to use their magnetic field maps to derive abundance distribution of Ca, however, due to limitations of the observational data they could not derive results of any significance. Our unique observational data set opens new possibilities to investigate the abundance distribution for Ca.

This element was studied by \citet{Ryabchikova97p1137}, and was used by \citet{Shulyak09p879} for their vertical stratification analysis, which showed that Ca has a vertical stratification profile that is similar to Fe. It is of particular interest to see how our magnetic field maps reproduce the spectral features of a chemical element with previously unknown abundance distribution. In the case of HD\,24712 the Ca lines are well suited for this purpose.

The line list of Ca lines, shown in Table~\ref{tbl:mdi:lines_ca}, was composed on basis of the same principles as discussed in Sect.~\ref{sec:mdi:lines} --- the Ca lines were selected based on their phase variations, presence of polarization signatures, and absence of blending by other spectral lines. 

We used the magnetic field maps (Fig.~\ref{fig:mf}) from the simultaneous MDI study of Fe, \nd{iii}, and Na as fixed parameters and derived abundance distribution of Ca. The resulting map for Ca is illustrated in Fig.~\ref{fig:abn_ca}. The comparison of the computed and observed Ca line profiles is presented in Fig.~\ref{fig:prfs_ca}, which shows that most polarization signatures in the Stokes~$QUV$ profiles are reproduced satisfactorily. There are some exceptions in the Stokes~$I$ profiles for several lines in particular for \ion{Ca}{i}~6162.173\,\AA{} and \ion{Ca}{i}~5857.451\,\AA. We suspect that this is caused by an inhomogeneous vertical distribution of Ca in the atmosphere of HD\,24712.

The abundance map of Ca exhibits variations over the stellar surface between -5.4\,dex and -7.1\,dex. Interestingly, Ca shows the presence of an overabundance area \bfa{close to} the magnetic equator and a roughly circular patch at the visible magnetic pole where Ca is depleted relative to its median value. From the comparison of Fig.~\ref{fig:abn} and \ref{fig:abn_ca} we see that this coincides with the abundance map of Fe, which has similar circular patch around the visible magnetic pole, where Fe is underabundant. One should be careful when interpreting these results as some Ca and Fe lines can be influenced by vertical stratification. It is possible that the roughly circular underabundance patches at the location of the visible magnetic pole derived in our MDI inversions are an artifact caused by ignoring vertical stratification.

\section{Discussion}
\label{sec:dis}
\subsection{Theoretical interpretation of the magnetic field in HD\,24712}
The discovery of dipole-like, axisymmetric magnetic field for HD\,24712 from our MDI study of four Stokes parameter observations provides a new perspective on the question of the origin and stability of magnetic fields in intermediate-mass main sequence A and late B stars. The general idea behind the existence of magnetic fields in these stars is the so-called "fossil field" hypothesis, which suggests that the magnetic fields of these stars are remnants from an earlier evolutionary phase. This hypothesis is supported by the long-term stability and large-scale nature of these magnetic fields coupled with the wide range of field strengths observed. One of the challenges of this theory was to find stable magnetic field equilibrium configurations that can be present in stellar interiors. Recent theoretical advances \citep{Braithwaite06p1077,Braithwaite2008p1947,Braithwaite2009p763} have shown that stable magnetic fields can exist in the interiors of these stars. \bfa{\citet{Braithwaite06p1077,Braithwaite2008p1947} found that for a stable, stratified star with initial random magnetic field configuration, its magnetic field evolves on Alfve\'n time-scales (tens of years for the typical Ap stars) into a stable equilibrium configuration of twisted flux tubes.}

These equilibrium configurations appear to depend on the initial conditions \citep{Braithwaite2009p763}, which may explain the large variety of magnetic field configurations inferred from MDI studies. It was discovered that there are roughly two characteristic stable configurations --- an approximately axisymmetric equilibrium with one flux tube forming a circle around the equator, and a non-axisymmetric equilibrium solution with one or more flux tubes arranged in a more complex pattern. The axisymmetric configuration in observations presents itself roughly as a dipole, with smaller contributions from higher multipoles. In contrast to that, the non-axisymmetric equilibrium configurations result in a highly complex magnetic field geometries such as the ones found for $\tau$\,Sco \citep{Donati06p629} and HD\,37776 \citep{Kochukhov2011p24}. 

An interesting conjecture proposed by \citet{Braithwaite06p1077} is that Ap stars with near-exact dipole fields should be older than Ap stars with more structure on smaller scales. This conjecture can be verified only by four Stokes parameters MDI studies. 

At this stage only two stars in addition to the analysis of HD\,24712 presented here, have been investigated with the MDI technique using all Stokes parameters. \citet{Kochukhov10p13} studied $\alpha^2$\,CVn and found dipole-like magnetic field with small-scale features, which have much higher field strength than in the surrounding areas. The second object is 53\,Cam for which \citet{Kochukhov2004p613} showed that quasi-dipolar magnetic field with strong toroidal components at angular scales of $30^\circ\text{--}\,40^\circ$ is necessary in order to fit the available spectropolarimetric observations.

For the purpose of this analysis we used the stellar ages of our objects published in the paper by \citet{Kochukhov2006p763}; 9.07 for HD\,24712, 8.27 for $\alpha^2$\,CVn, and 8.84 for 53~Cam. The values adopted here are in $\log t$\,(yr) units. 

Given these results, our finding of dipole-like, axisymmetric magnetic field for HD\,24712, which is the oldest star in our three star sample, gives some evidence for the hypothesis that old Ap stars have predominantly dipolar magnetic fields with little structure on small scales.

Our future plans involve performing MDI in four Stokes parameters of other Ap/Bp stars with different masses and ages, which will help us further assess this hypothesis and investigate the dependence of the magnetic field geometry on other stellar parameters.

\subsection{Abundance distribution of chemical elements and atomic diffusion}
It is believed that the patchy abundance distribution and vertical stratification of chemical elements in the atmospheres of Ap stars is caused by atomic diffusion in the presence of magnetic field \citep{Michaud1981p244,LeBlanc2009,Alecian2010p53}. According to this model some elements are expected to accumulate and form cloud-like structures in the atmospheres of Ap stars. However, theoretical predictions arising from this theory have in general been only qualitative in the case of magnetic Ap stars, with little predictive power for individual stars.

\bfa{We} compare our empirical abundance maps from the latest MDI analysis of HD\,24712 to the theoretical results of \citet{Alecian2010p53}, who computed 2-dimensional stratification for a number of elements for several values of effective temperature and dipole field strength. \bfa{These authors deduced the appearance of narrow belts of enhanced metals around the magnetic equator for stars with dipole magnetic field. For stars with non-dipolar field the abundance patches would be created at places where the field lines are parallel to the local surface.} The abundance maps for Fe, Nd, and Na derived in this study do not show such abundance structures. Instead, we see that the Fe and Ca abundance maps very roughly correlate with the field modulus (compare Fig.~\ref{fig:abn} and \ref{fig:abn_ca} to Fig.~\ref{fig:mf} for Fe and Ca). The abundance maps for Nd and Na anti-correlate between each other and show strong abundance gradients between the two polar regions. \citet{Alecian2010p53} suggested that such large abundance difference between the polar regions might be caused by non-dipolar configuration, or decentered dipole with significant difference in field strength at the poles. It appears that our results from MDI inversion using four Stokes vector observations of HD\,24712 refute this suggestion as the newly derived magnetic field is dipole-like to a very high extent, and although the magnetic field maps do indeed show slight asymmetry of approximately 0.6\,kG between the poles we do not consider this to be ''significant`` difference in strength. 

We conclude that our abundance maps of Fe, Nd, Na, and Ca do not show behavior consistent with the presence of \bfa{narrow} overabundance belt \bfa{around the magnetic equator}. At this point, we believe that better theoretical models are required in order to explain the variety of observed abundance distributions of chemical elements in the atmosphere of HD\,24712. 

\section{Conclusions}
\label{sec:concl}
We presented results of the MDI study of the cool Ap star HD\,24712. This is first such analysis for a rapidly oscillating Ap star performed on the basis of phase-resolved spectropolarimetric observations of line profiles in all four Stokes parameters that have resolving power exceeding $10^5$ and signal-to-noise ratio of 300--600.

The main results from our investigation are summarized below.
\begin{itemize}
\item The magnetic field topology of HD\,24712 is mostly poloidal and dipolar with small contributions from higher order harmonics. We found a small difference between the positive and negative magnetic pole of approximately 0.6\,kG in field modulus. This can be attributed to the nature of the magnetic field configuration of HD\,24712, for which only the positive magnetic pole is visible by an observer from Earth.


\item \bfa{The recovered abundance distribution map of Fe shows enhancements on both sides of the magnetic equator. The abundance map of Ca shows enhancements only on the side of the magnetic equator with the positive magnetic pole. Both maps show presence of an underabundance patch centered at the visible magnetic pole.} The polar feature may represent an artifact caused by ignoring vertical chemical stratification of Fe and Ca in the atmosphere of HD\,24712.

\item The recovered abundance maps of Nd and Na show strong anti-correlation between them. Nd is highly abundant around the visible magnetic pole, while Na is overabundant around the negative magnetic pole.
\item We find tentative evidence for the hypothesis that Ap stars with dipole-like fields that have weak contribution from higher-order harmonics are older or less-massive than stars with magnetic fields that have more small-scale structures. 
\item The derived abundance maps of Fe, Nd, Na, and Ca are not consistent with current theoretical predictions of atomic diffusion theory in the presence of magnetic fields.
\end{itemize}

\begin{acknowledgements}
OK is a Royal Swedish Academy of Sciences Research Fellow, supported by grants from Knut and Alice Wallenberg Foundation and Swedish Research Council. The computations presented in this paper were performed on resources provided by SNIC through Uppsala Multidisciplinary Center for Advanced Computational Science (UPPMAX) under project snic2013-11-24. \bfa{TR acknowledges partial financial support from the Presidium RAS Program ``Nonstationary Phenomena in Objects of the Universe.''} Resources provided by the electronic databases (VALD, Simbad, NASA ADS) are gratefully acknowledged.
\end{acknowledgements}

\bibliographystyle{aa}
\bibliography{articles}

\begin{thebibliography}{38}
\expandafter\ifx\csname natexlab\endcsname\relax\def\natexlab#1{#1}\fi

\bibitem[{Alecian \& Stift(2010)}]{Alecian2010p53}
Alecian, G. \& Stift, M.~J. 2010, \aap, 516, A53

\bibitem[{Bagnulo {et~al.}(1995)Bagnulo, Landi~Degl'Innocenti, Landolfi, \&
  Leroy}]{Bagnulo1995p459}
Bagnulo, S., Landi~Degl'Innocenti, E., Landolfi, M., \& Leroy, J.~L. 1995,
  \aap, 295, 459

\bibitem[{{Bagnulo} {et~al.}(2001){Bagnulo}, {Wade}, {Donati}, {Landstreet},
  {Leone}, {Monin}, \& {Stift}}]{Bagnulo2001p889}
{Bagnulo}, S., {Wade}, G.~A., {Donati}, J.-F., {et~al.} 2001, \aap, 369, 889

\bibitem[{{Braithwaite}(2008)}]{Braithwaite2008p1947}
{Braithwaite}, J. 2008, \mnras, 386, 1947

\bibitem[{{Braithwaite}(2009)}]{Braithwaite2009p763}
{Braithwaite}, J. 2009, \mnras, 397, 763

\bibitem[{Braithwaite \& Nordlund(2006)}]{Braithwaite06p1077}
Braithwaite, J. \& Nordlund, {\AA}. 2006, \aap, 450, 1077

\bibitem[{{Donati} \& {Brown}(1997)}]{Donati1997p1135}
{Donati}, J.-F. \& {Brown}, S.~F. 1997, \aap, 326, 1135

\bibitem[{Donati {et~al.}(2006)Donati, Howarth, Jardine, Petit, Catala,
  Landstreet, Bouret, Alecian, Barnes, Forveille, Paletou, \&
  Manset}]{Donati06p629}
Donati, J.-F., Howarth, I.~D., Jardine, M.~M., {et~al.} 2006, \mnras, 370, 629

\bibitem[{Kochukhov \& Bagnulo(2006)}]{Kochukhov2006p763}
Kochukhov, O. \& Bagnulo, S. 2006, \aap, 450, 763

\bibitem[{{Kochukhov} {et~al.}(2004){Kochukhov}, {Bagnulo}, {Wade}, {Sangalli},
  {Piskunov}, {Landstreet}, {Petit}, \& {Sigut}}]{Kochukhov2004p613}
{Kochukhov}, O., {Bagnulo}, S., {Wade}, G.~A., {et~al.} 2004, \aap, 414, 613

\bibitem[{{Kochukhov} {et~al.}(2014){Kochukhov}, {L{\"u}ftinger}, {Neiner},
  {Alecian}, \& {MiMeS Collaboration}}]{Kochukhov2014p83}
{Kochukhov}, O., {L{\"u}ftinger}, T., {Neiner}, C., {Alecian}, E., \& {MiMeS
  Collaboration}. 2014, \aap, 565, A83

\bibitem[{{Kochukhov} {et~al.}(2011){Kochukhov}, {Lundin}, {Romanyuk}, \&
  {Kudryavtsev}}]{Kochukhov2011p24}
{Kochukhov}, O., {Lundin}, A., {Romanyuk}, I., \& {Kudryavtsev}, D. 2011, \apj,
  726, 24

\bibitem[{Kochukhov \& Piskunov(2002)}]{Kochukhov2002p868}
Kochukhov, O. \& Piskunov, N. 2002, \aap, 388, 868

\bibitem[{Kochukhov {et~al.}(2004)Kochukhov, Ryabchikova, \&
  Piskunov}]{Kochukhov04p13}
Kochukhov, O., Ryabchikova, T., \& Piskunov, N. 2004, \aap, 415, L13

\bibitem[{Kochukhov \& Wade(2010)}]{Kochukhov10p13}
Kochukhov, O. \& Wade, G.~A. 2010, \aap, 513, A13

\bibitem[{{Kochukhov} {et~al.}(2012){Kochukhov}, {Wade}, \&
  {Shulyak}}]{Kochukhov2012p3004}
{Kochukhov}, O., {Wade}, G.~A., \& {Shulyak}, D. 2012, \mnras, 421, 3004

\bibitem[{Kupka {et~al.}(1999)Kupka, Piskunov, Ryabchikova, Stempels, \&
  Weiss}]{Kupka1999p119}
Kupka, F., Piskunov, N., Ryabchikova, T.~A., Stempels, H.~C., \& Weiss, W.~W.
  1999, \aaps, 138, 119

\bibitem[{Landolfi {et~al.}(1993)Landolfi, Landi~Degl'Innocenti,
  Landi~Degl'Innocenti, \& Leroy}]{Landolfi1993p285}
Landolfi, M., Landi~Degl'Innocenti, E., Landi~Degl'Innocenti, M., \& Leroy,
  J.~L. 1993, \aap, 272, 285

\bibitem[{Landstreet \& Mathys(2000)}]{Landstreet2000p213}
Landstreet, J.~D. \& Mathys, G. 2000, \aap, 359, 213

\bibitem[{{LeBlanc} {et~al.}(2009){LeBlanc}, {Monin}, {Hui-Bon-Hoa}, \&
  {Hauschildt}}]{LeBlanc2009}
{LeBlanc}, F., {Monin}, D., {Hui-Bon-Hoa}, A., \& {Hauschildt}, P.~H. 2009,
  \aap, 495, 937

\bibitem[{Leroy(1995)}]{Leroy1995p79}
Leroy, J.~L. 1995, \aaps, 114, 79

\bibitem[{L{\"u}ftinger {et~al.}(2010)L{\"u}ftinger, Kochukhov, Ryabchikova,
  Piskunov, Weiss, \& Ilyin}]{Lueftinger10p71}
L{\"u}ftinger, T., Kochukhov, O., Ryabchikova, T., {et~al.} 2010, \aap, 509,
  A71+

\bibitem[{{Mashonkina} {et~al.}(2005){Mashonkina}, {Ryabchikova}, \&
  {Ryabtsev}}]{Mashonkina2005p309}
{Mashonkina}, L., {Ryabchikova}, T., \& {Ryabtsev}, A. 2005, \aap, 441, 309

\bibitem[{Mathys(1988)}]{Mathys1988p179}
Mathys, G. 1988, \aap, 189, 179

\bibitem[{Mathys \& Hubrig(1997)}]{Mathys97p475}
Mathys, G. \& Hubrig, S. 1997, \aaps, 124, 475

\bibitem[{Mayor {et~al.}(2003)Mayor, Pepe, Queloz, Bouchy, Rupprecht, Lo~Curto,
  Avila, Benz, Bertaux, Bonfils, Dall, Dekker, Delabre, Eckert, Fleury,
  Gilliotte, Gojak, Guzman, Kohler, Lizon, Longinotti, Lovis, Megevand,
  Pasquini, Reyes, Sivan, Sosnowska, Soto, Udry, van Kesteren, Weber, \&
  Weilenmann}]{Mayor2003p20}
Mayor, M., Pepe, F., Queloz, D., {et~al.} 2003, The Messenger, 114, 20

\bibitem[{{Michaud} {et~al.}(1981){Michaud}, {Charland}, \&
  {Megessier}}]{Michaud1981p244}
{Michaud}, G., {Charland}, Y., \& {Megessier}, C. 1981, \aap, 103, 244

\bibitem[{Piskunov \& Kochukhov(2002)}]{Piskunov2002p736}
Piskunov, N. \& Kochukhov, O. 2002, \aap, 381, 736

\bibitem[{Piskunov {et~al.}(2011)Piskunov, Snik, Dolgopolov, Kochukhov,
  Rodenhuis, Valenti, Jeffers, Makaganiuk, Johns-Krull, Stempels, \&
  Keller}]{Piskunov11p7}
Piskunov, N., Snik, F., Dolgopolov, A., {et~al.} 2011, The Messenger, 143, 7

\bibitem[{Rusomarov {et~al.}(2013)Rusomarov, Kochukhov, Piskunov, Jeffers,
  Johns-Krull, Keller, Makaganiuk, Rodenhuis, Snik, Stempels, \&
  Valenti}]{Rusomarov2013p8}
Rusomarov, N., Kochukhov, O., Piskunov, N., {et~al.} 2013, \aap, 558, A8

\bibitem[{{Ryabchikova} {et~al.}(2007){Ryabchikova}, {Sachkov}, {Kochukhov}, \&
  {Lyashko}}]{Ryabchikova2007p907}
{Ryabchikova}, T., {Sachkov}, M., {Kochukhov}, O., \& {Lyashko}, D. 2007, \aap,
  473, 907

\bibitem[{Ryabchikova {et~al.}(1997)Ryabchikova, Landstreet, Gelbmann, Bolgova,
  Tsymbal, \& Weiss}]{Ryabchikova97p1137}
Ryabchikova, T.~A., Landstreet, J.~D., Gelbmann, M.~J., {et~al.} 1997, \aap,
  327, 1137

\bibitem[{Shulyak {et~al.}(2009)Shulyak, Ryabchikova, Mashonkina, \&
  Kochukhov}]{Shulyak09p879}
Shulyak, D., Ryabchikova, T., Mashonkina, L., \& Kochukhov, O. 2009, \aap, 499,
  879

\bibitem[{{Silvester} {et~al.}(2014){Silvester}, {Kochukhov}, \&
  {Wade}}]{Silvester2014p182}
{Silvester}, J., {Kochukhov}, O., \& {Wade}, G.~A. 2014, \mnras, 440, 182

\bibitem[{{Silvester} {et~al.}(2012){Silvester}, {Wade}, {Kochukhov},
  {Bagnulo}, {Folsom}, \& {Hanes}}]{Silvester2012p1003}
{Silvester}, J., {Wade}, G.~A., {Kochukhov}, O., {et~al.} 2012, \mnras, 426,
  1003

\bibitem[{Snik {et~al.}(2011)Snik, Kochukhov, Piskunov, Rodenhuis, Jeffers,
  Keller, Dolgopolov, Stempels, Makaganiuk, Valenti, \&
  Johns-Krull}]{Snik2011p237}
Snik, F., Kochukhov, O., Piskunov, N., {et~al.} 2011, in Astronomical Society
  of the Pacific Conference Series, Vol. 437, Solar Polarization 6, ed. J.~R.
  Kuhn, D.~M. Harrington, H.~Lin, S.~V. Berdyugina, J.~Trujillo-Bueno, S.~L.
  Keil, \& T.~Rimmele, 237

\bibitem[{Stibbs(1950)}]{Stibbs1950p395}
Stibbs, D.~W.~N. 1950, \mnras, 110, 395

\bibitem[{Wade {et~al.}(2000)Wade, Donati, Landstreet, \& Shorlin}]{Wade00p823}
Wade, G.~A., Donati, J.-F., Landstreet, J.~D., \& Shorlin, S.~L.~S. 2000,
  \mnras, 313, 823

\end{thebibliography}


\end{document}